\documentclass[prb,aps,amsmath,amssymb,amsfonts,floatfix,superscriptaddress,showpacs,nofootinbib,twocolumn]{revtex4-1}
\usepackage{graphicx}
\usepackage{color}
\usepackage[utf8]{inputenc}
\usepackage{dcolumn}
\usepackage{hyperref}
\usepackage[caption=false]{subfig}
\usepackage[english]{babel}
\usepackage{indentfirst}
\hypersetup{colorlinks=true,breaklinks,linkcolor=blue,urlcolor=blue,citecolor=blue}

\usepackage{tikz}
\usetikzlibrary{calc}
\usetikzlibrary{fit}
\usetikzlibrary{decorations.pathmorphing}
\usetikzlibrary{decorations.pathreplacing}
\usetikzlibrary{decorations.markings}
\usetikzlibrary{arrows}

\newcommand{\av}[1]{\ensuremath{\left\langle #1 \right\rangle}}
\newcommand{\up}{\uparrow}
\newcommand{\dn}{\downarrow}

\newcommand{\qv}{\mathbf{q}}
\newcommand{\kv}{\mathbf{k}}

\newcommand{\sz}{\text{sz}}
\newcommand{\ch}{\text{ch}}

\DeclareMathOperator{\abs}{abs}

\begin{document}

\title{
Dual Boson approach with instantaneous interaction
}

\author{L. Peters}
%\email{L.Peters@science.ru.nl}
\affiliation{Radboud University, Institute for Molecules and Materials, 6525AJ Nijmegen, The Netherlands}

\author{E. G. C. P. van Loon}
\affiliation{Radboud University, Institute for Molecules and Materials, 6525AJ Nijmegen, The Netherlands}
\affiliation{Institut f{\"u}r Theoretische Physik, Universit{\"a}t Bremen, Otto-Hahn-Allee 1, 28359 Bremen, Germany}
\affiliation{Bremen Center for Computational Materials Science, Universit{\"a}t Bremen, Am Fallturm 1a, 28359 Bremen, Germany}

\author{A. N. Rubtsov}
\affiliation{Russian Quantum Center, 143025 Skolkovo,  Russia}
\affiliation{Department of Physics, M.V. Lomonosov Moscow State University, 119991 Moscow, Russia}

\author{A. I. Lichtenstein}
\affiliation{Institute of Theoretical Physics, University of Hamburg, 20355 Hamburg, Germany}

\author{M. I. Katsnelson}
\affiliation{Radboud University, Institute for Molecules and Materials, 6525AJ Nijmegen, The Netherlands}

\author{E. A. Stepanov}
\affiliation{Institute of Theoretical Physics, University of Hamburg, 20355 Hamburg, Germany}
\affiliation{Radboud University, Institute for Molecules and Materials, 6525AJ Nijmegen, The Netherlands}
%\affiliation{Centre de Physique Th\'eorique, Ecole Polytechnique, CNRS UMR 7644, 91128 Palaiseau, France}

\begin{abstract}
The Dual Boson approach to strongly correlated systems generally involves a dynamic (frequency-dependent) interaction in the auxiliary impurity model. In this work, we explore the consequences of forcing this interaction to be instantaneous (frequency-independent) via the use of a self-consistency condition on the instantaneous susceptibility. 
The result is a substantial simplification of the impurity model, especially with an eye on realistic multiband implementations, while keeping desireable properties of the Dual Boson approach, such as the charge conservation law, intact. 
We show and illustrate numerically that this condition enforces the absence of phase transitions in finite systems, as should be expected from general physical considerations, and respects the Mermin-Wagner theorem. In particular, the theory does not allow the metal to insulator phase transition associated with the formation of the magnetic order in a two-dimensional system. At the same time, the metal to charge ordered phase transition is allowed, as it is not associated with the spontaneous breaking of a continuous symmetry, and is accurately captured by the introduced approach.
\end{abstract}

\maketitle

\section{Introduction}

The Hubbard model~\cite{Hubbard63,Kanamori63,Gutzwiller63,Hubbard64} describes interacting itinerant electrons. 
The Coulomb interaction between these electrons leads to correlations and makes this a many-body problem. 
When interaction and itinerancy are almost equally strong, the resulting electronic correlations make the system notoriously hard to study. 

The Dynamical Mean-Field Theory (DMFT)~\cite{Georges96} is an important tool to describe correlated electrons. 
The main idea of the method is to use an auxiliary ``impurity'' model that consists of only a single site and contains the most important correlations of the full system.
This approach is exact in the limit of infinite dimension~\cite{Metzner89}, in finite dimensional systems it still serves as a useful approximation. 

Extensions of DMFT have been developed~\cite{Rohringer18} to add the spatial correlations and non-local interactions that DMFT ignores. 
These extensions add additional correlation effects on top of DMFT, but they can also change the impurity model that is used as a starting point.
To account for screening by the non-local Coulomb interaction, Extended DMFT (EDMFT)~\cite{Sengupta95,Si96,Kajueter96,Smith00,Chitra00,Chitra01} and its diagrammatic extensions~\cite{Rohringer18}, introduce dynamic interactions into the impurity model via an effective frequency dependent bosonic hybridization function. While the EDMFT still considers electronic correlations at the level of the local impurity problem, the Dual Boson (DB) theory~\cite{Rubtsov12, Stepanov16} and other EDMFT-based approaches take additional nonlocal correlation effects into account diagrammatically.

The DB approach aims to treat fermionic and collective bosonic degrees of freedom on an equal footing. 
Nevertheless, until recently the DB calculations were mostly focused on the description of charge degrees of freedom, with the bosonic hybridization function introduced only in the charge channel. This means that the screening by collective spin fluctuations was missing. The introduction of dynamic interactions in the spin channel generally comes at a computational cost~\cite{Otsuki13,Steiner15}.
For single-band systems, a dynamic spin density ($S^z$-$S^z$) interaction
can be incorporated into CT-QMC impurity solvers at moderate cost, in the same way as the dynamic charge interaction~\cite{Werner07,Werner10}. However, this immediately breaks the spin-rotational invariance of the local impurity problem. On the other hand, the use of the same frequency dependent bosonic hybridization function for every ($x$, $y$, and $z$) spin channel to obey the rotational invariance immediately leads to a violation of conservation laws~\cite{Krien17}.

To address these situations, a simplification of the Dual Boson approach that does not require dynamic interactions in the impurity model has been introduced recently~\cite{Stepanov18}.
We use the fact that the dynamic interactions in Dual Boson are \emph{a priori} free parameters in the Hubbard-Stratonovich decoupling that leads to the impurity model. 
The dynamic interaction is usually determined using a set of self-consistency conditions, one for every frequency. 
By forcing the interaction to be instantaneous (independent of frequency), the number of free parameters is drastically reduced, with the benefit of having a much simpler impurity model that accounts for the screening by collective charge and spin fluctuations in a spin-rotational invariant form that does not violate local conservation laws. From the physical point of view, the constant form of the hybridization function in the spin channel can be motivated by the fact that collective spin fluctuations are slower and have lower energy than single-particle (electronic) excitations. Thus the interaction between spins and electrons is instantaneous ($\delta$-function in time), which leads to a constant hybridization function.

So far, this method has not been systematically tested. In this work we aim to systematically test its performance in the description of collective electronic instabilities of the two- and three-dimensional extended Hubbard model.
We compare this instantaneous interaction Dual Boson approach with the traditional dynamic interaction Dual Boson approach and with the Dynamical Mean-Field which always keeps the interaction of the original Hubbard model.
We also discuss the relationship with the two-particle self-consistent approach~\cite{Vilk94,Vilk97} and the Moriya correction in D$\Gamma$A~\cite{Toschi07}, which are also based on self-consistently renormalizing an effective interaction.

\section{Model and Method}

We consider the half-filled extended Hubbard model on the square and cubic lattice,
\begin{align}
 H =& -  \sum_{i,j,\sigma}t^{\phantom{\dagger}}_{ij} \, c^\dagger_{j\sigma} c^{\phantom{\dagger}}_{i\sigma} + U \sum_i n_{i\up} n_{i\dn} \notag \\
 &+ \frac{1}{2} \sum_{i,j} V_{ij} \left(n_{i}-\av{n_i}\right) \left(n_{j}-\av{n_j}\right).
\end{align}
Here $c^\dagger_{i\sigma}$ and $c^{\phantom{\dagger}}_{i\sigma}$ are creation and annihilation operators for an electron on site $i$ with spin $\sigma$ and $n_{i\sigma} = c^\dagger_{i\sigma} c^{\phantom{\dagger}}_{i\sigma}$ is the corresponding number operator. 
The total electronic density on site $i$ is equal to $n_i=n_{i\up}+n_{i\dn}$. 
The physical parameters of the model are the hopping amplitude $t_{ij}$, the on-site Coulomb interaction $U$ and the intersite Coulomb interaction  $V_{ij}$.
We consider only nearest-neighbour hopping and interaction, $t_{ij}=t$ and $V_{ij}=V$ when $i$ and $j$ are nearest-neighbors and 0 otherwise.

For $V=0$, this model is simply the Hubbard model.
In this work, we consider the (extended) Hubbard model at half-filling, $\av{n_i}=1$. This is obtained by setting $\mu=U/2$.
We restrict ourselves to phases without explicit symmetry breaking. We do calculate the susceptibility associated with antiferromagnetic (AF) and charge density wave (CDW) phases to check for instabilities towards ordered phases.

The idea of the Dual Fermion~\cite{Rubtsov08} and Dual Boson~\cite{Rubtsov12} methods is to decouple the extended Hubbard model into two parts. 
The first is an impurity part that only contains local degrees of freedom.
This part can be solved numerically exactly. 
The impurity part should be chosen in such a way that it contains the most important correlation effects, since all correlations that are present in the impurity model are treated exactly.
The second part is the remainder, which is a lattice model just like the original (extended) Hubbard model. 
The difference with the original model is that the original degrees of freedom are transformed to new \emph{dual} degrees of freedom when the impurity model is integrated out.
These dual degrees of freedom are less correlated, so that the correlations in this dual part can be addressed perturbatively.

Here, we provide an overview of the formulation of the Dual Boson method, for a full derivation and analysis of the method we refer the reader to the original works~\cite{Rubtsov12,vanLoon14,Hafermann14b,Stepanov16} and the review~\cite{Rohringer18}.
Mathematically, the action formalism is the most convenient way to perform the dual decoupling. 
We write the action in terms of the Fourier transforms of the hopping and interaction, $t_\kv$ and $V_\qv$, and in terms of Matsubara frequencies
\begin{align}
{\cal S} =& -\sum_{i,\nu,\sigma} c^{*}_{i\nu\sigma}[i\nu+\mu]c_{i\nu\sigma} + U\sum_{i,\omega}n_{i,\omega\uparrow}n_{i,-\omega\downarrow}\notag\\
&+ \sum_{\kv,\nu,\sigma}t^{\phantom{*}}_{\kv} \, c^{*}_{\kv\nu\sigma}c^{\phantom{*}}_{\kv\nu\sigma} + \frac{1}{2}\sum_{\qv,\omega}V_{\qv} \, n_{\qv\omega}n_{-\qv-\omega} \\
 =& \sum_i {\cal S}_\text{imp} + {\cal S}_{\text{remainder}}.
\end{align}
This separation is made by introducing fermionic and bosonic hybridization functions $\Delta$, $\Lambda^{\rho}$, where $\rho$ is a bosonic channel, e.g., charge or $S^z$. 
\begin{align}
{\cal S}_{\text{imp}} =& -\sum_{\nu,\sigma} c^{*}_{\nu\sigma}[i\nu+\mu-\Delta^{\phantom{*}}_{\nu}]c^{\phantom{*}}_{\nu\sigma}\notag\\
&+ U\sum_{\omega}n_{\omega\uparrow}n_{-\omega\downarrow} + \frac{1}{2}\sum_{\rho,\omega} \Lambda^{\rho}_{\omega} \, \rho_{\omega} \rho_{-\omega}.
\end{align}
These hybridizations $\Delta$, $\Lambda$ can be chosen freely. Below in Section~\ref{sec:sc} we discuss how to make this choice.
In general, the Dual Boson approach considers $\Lambda$ to be a function of $\omega$. 
In this work, we restrict ourselves to constant $\Lambda$. 
In that case, the impurity action can simply be rewritten as
\begin{align}
{\cal S}_{\text{imp}} = & -\sum_{\nu,\sigma} c^{*}_{\nu\sigma}[i\nu+\mu-\Delta^{\phantom{*}}_{\nu}]c^{\phantom{*}}_{\nu\sigma}+ U' \sum_{\omega}n_{\omega\up} n_{-\omega\dn},
\end{align}
where $U'=U+\Lambda_{\text{ch}}-\Lambda_{\text{sz}}$.
For the impurity model to also be at half-filling, we take $\mu'=U'/2$.
This choice of hybridization functions is useful, since the resulting form of the impurity action does not introduce higher-order terms in the Ward identities~\cite{Krien17,Stepanov18}. As we discuss below in Section~\ref{sec:implementation}, the impurity problem is solved numerically exactly and provides full frequency dependent local quantities such as the one- and two-particle Green's functions and fermion-fermion and fermion-boson vertices needed for the construction of the dual diagrams.

The spatial correlations in the dual part of the problem are addressed using diagrammatic methods. 
In this work, we concentrate on the correlation effects in the susceptibility and we do not calculate dual self-energy corrections.
The following definition for the susceptibility $X_{\qv\omega} = -\av{\rho \rho}_{\qv\omega}$ can be applied both to the charge channel, with $\rho=n-\av{n}$, and to the spin channel, with $\rho=S^z-\av{S^z}$. 
The Dual Boson expression is given by
\begin{align}
 X^{-1}_{\qv\omega} =& \mathcal{X}^{-1}_{\qv\omega} + \Lambda_\omega - V_\qv \label{eq:X},
\end{align}
where
\begin{align}
 \mathcal{X}_{\qv\omega} =& \chi_\omega + \chi_\omega \tilde{\Pi}_{\qv\omega} \chi_\omega,
\end{align}
and
$\chi_\omega = -\av{\rho \rho}^\text{imp}_\omega$ is the corresponding  susceptibility of the impurity model. 
The non-local correlation effects are contained in $\tilde{\Pi}$, the dual polarization operator. In this work, we use the ladder equation to calculate $\tilde{\Pi}$~\cite{Rubtsov12,vanLoon14,Hafermann14b,Stepanov16}, which describes repeated particle-hole scattering mediated by the fermion-fermion vertex of the impurity model. The ladder DB approach with the frequency dependent bosonic hybridization shows a quantitatively good result for the susceptibility in agreement with Quantum Monte Carlo and Dynamical Cluster Approximation calculations in a broad range of Coulomb interactions and dopings~\cite{PhysRevB.94.165141, PhysRevB.95.115149, PhysRevB.99.115124}. The restriction to instantaneous interactions simply leads to the replacement of $\Lambda_\omega$ by a constant $\Lambda$ in Eq.~\eqref{eq:X}.

Dynamical Mean-Field Theory usually deals with the Hubbard model, with $V=0$. 
The DMFT susceptibility is obtained by setting $\Lambda=0$ in Eq.~\eqref{eq:X}. 
This formulation of the DMFT susceptibility is equivalent~\cite{Hafermann14b} to more traditional ways of computing it~\cite{Georges96}. 
The extended Hubbard model's intersite Coulomb interactions can be included in the DMFT susceptibility via Eq.~\eqref{eq:X} with $\Lambda=0$, this essentially corresponds to an RPA-like treatment of these additional interactions, in which they do not alter the impurity or the single-particle properties.

\subsection{Self-consistency condition}
\label{sec:sc}

So far, we have not specified the interaction renormalization $\Lambda$, which is determined by a self-consistency condition.
We use the ``lattice'' self-consistency condition proposed for the self-consistent Dual Boson approach~\cite{Stepanov16}, except that here we have only two free parameters $\Lambda^{\text{ch}}$, $\Lambda^{\text{sz}}$ instead of two dynamic functions of frequency.
This means we only need one self-consistency condition per channel instead of having a self-consistency for every Matsubara frequency. 
\begin{align}
 \sum_{\omega} \chi^{\text{ch/sz}}_{\omega} = \sum_{\omega} X^{\text{ch/sz}}_{\text{loc, }\omega},
\end{align}
where $X_{\text{loc}}$ is the local part of the lattice susceptibility, obtained as the average over momenta of $X_{\qv}$. As shown in Ref.~\onlinecite{Stepanov18}, this choice of the self-consistency condition follows from the invariance of the initial lattice problem with respect to the variation of the introduced hybridization functions, and it fulfills the Pauli principle.

The sum over frequencies in the self-consistency condition corresponds to taking the equal-time component of the susceptibility\footnote{The equal time component is obtained by dividing both sides by $\beta$ to normalize the frequency sum, and the sum runs over both positive and negative Matsubara frequencies.}, so we can write
\begin{align}
 \chi^{\text{ch/sz}}(\tau=0) = X^{\text{ch/sz}}_{\text{loc}}(\tau=0). \label{eq:sc:time}
\end{align}

The difference between the local, equal time charge and spin susceptibility determines the double occupancy $D=\av{n_\up n_\dn}$.
The self-consistency condition used here assures that the double occupancy of the lattice susceptibility is equal to the one of the impurity model~\cite{vanLoon16}.
On the other hand, the potential energies will be different since the lattice model and the impurity model have different interactions $U$ and $U'$.
This type of inconsistency between impurity and lattice generally occurs in approximate methods based on DMFT~\cite{Krien17}.
Note that the argument about potential energies applies to the Hubbard model ($V=0$), for the extended Hubbard model there is no direct correspondence between the potential energy of the impurity and the lattice since the latter also contains intersite contributions.

Dual Boson calculations with dynamic interactions in the $S^z$ channel break the spin rotational symmetry of the impurity model if they do not have the same dynamic interaction in the $S^x$ and $S^y$ channels. There is no such rotational symmetry breaking in the current approach, since the impurity model has the rotationally invariant interaction $U'$.
In the dual part of the calculations, it is sufficient to calculate only the $S^z$ channel since the $S^x$ and $S^y$ channel follow from rotational symmetry.

Appendix~\ref{app:analytics} describes analytical results for the self-consistency condition in several simplifying limits. These results carry over from the self-consistent Dual Boson approach with dynamic interactions~\cite{Stepanov16}. 

The reader might wonder why the self-consistency condition is important in the first place. 
The relation between the original and the dual theory is exact for any choice of $\Delta$ and $\Lambda$.
However, the dual theory is only solved approximately, perturbatively. 
The choice of $\Lambda$ becomes important in this approximate theory.
In this sense, there is some similarity with the Fierz ambiguity~\cite{Baier00,Jaeckel03,Ayral17}, where the decomposition of $U$ into channels determines the computational outcome of approximate methods even though the exact system is unchanged by the choice of decomposition.
The present situation can also be seen through this lense, since the impurity part of the calculation depends only on the combination $U'=U+\Lambda_{\text{ch}}-\Lambda_{\text{sz}}$, whereas the dual part of the calculation and the transformation between dual and lattice quantities are affected by how $U'$ is decomposed into the charge and magnetic channel. 

As a corollary, the choice of self-consistency condition should be informed by the \emph{approximations} made in the dual theory, since only these approximations make the self-consistency condition relevant. Presently, the approximations are the elimination of vertices beyond the two-particle level, the restriction to ladder diagrams and finally the choice of the density and magnetic channels for the ladder.
Starting with the last point, the choice of channels for the ladder and subsequently for the susceptibility directly informs the choice of $\Lambda_{\text{ch}}$ and $\Lambda_{\text{sz}}$ as the self-consistent parameters (see, e.g., Ref.~\onlinecite{delRe18} for a recent discussion on the role of channels in the attractive Hubbard model).
Diagrammatic Monte Carlo approaches to the dual theory~\cite{Iskakov16,Gukelberger2017} do not restrict themselves to ladder diagrams and are therefore not restricted by the second and third point.

\subsection{Implementation}
\label{sec:implementation}

Apart from the self-consistency condition and the effective interaction, our method follows the Dual Boson approach~\cite{Rubtsov12} and its implementation~\cite{vanLoon14}. For the impurity model, we use a modified version of the open source CT-HYB solver~\cite{Hafermann13,Hafermann14} based on the ALPS libraries~\cite{ALPS2}.

The implementation of the Dual Boson approach consists of two computationally heavy parts, the impurity solver and the evaluation of the dual diagrams, both implemented in \texttt{C++} and linked together by a lighter \texttt{python} interface. 
The self-consistency condition is based entirely on the impurity susceptibility $\chi$ and the local part of the lattice susceptibility $X_{\text{loc}}$. 
These are part of the usual output of the impurity solver and the dual program.
This means that the self-consistency condition can be implemented in the \texttt{python} interface, the renormalized effective interaction $U'=U+\Lambda_{\text{ch}}-\Lambda_{\text{sz}}$ enters the impurity solver as a parameter and no other changes are needed.

We use an update formula with damping parameter $\xi$,
\begin{align}
\Lambda_{\text{new}} =  \Lambda_{\text{old}} + \xi \, \frac{X_{\text{loc}}(\tau=0)-\chi(\tau=0)}{X_{\text{loc}}(\tau=0) \chi(\tau=0)}.
\end{align}
A smaller value of $\xi$ makes the self-consistency procedure more stable at the cost of slower convergence, we take $\xi\in [0.1,1]$.
Similar to the self-consistent procedure to determine $U(\omega)$ used in Ref.~\onlinecite{vanLoon16}, we find that convergence slows down substantially at larger interaction strengths.

\subsection{Phase transitions and the Mermin-Wagner theorem}
\label{sec:merminwagner}

The goal of extensions of DMFT is to add the physical aspects of finite (low) dimensional physics to the $d=\infty$ solution. Dimensionality is especially important for the understanding of phase transitions to ordered phases. These show up as a divergence in the corresponding susceptibility, at zero frequency and at a specific momentum $q^\ast$ which describes the ordering pattern, e.g., $\qv_{\text{AF, 2d}}=(\pi,\pi)$ or $\qv_{\text{AF, 3d}}=(\pi,\pi,\pi)$ for antiferromagnetic ordering.

There is a natural link between phase transitions and the self-consistency condition when the susceptibility appears in the latter~\cite{Vilk97}.
We have previously written the self-consistency condition as
\begin{align}
 \sum_{\omega} \chi^{\text{sz}}_{\omega} = \frac{1}{N} \sum_{\omega} \sum_\qv X^{\text{sz}}_{\qv\omega}. \label{eq:sc:divergence}
\end{align}
Here, the momentum sum signifies a Brillouin Zone integral in the case of an infinitely large system.
A phase transition shows up as a divergence in $X_{\qv^\ast,\omega=0}$ on the right-hand side of Eq.~\eqref{eq:sc:divergence}.
However, the left-hand side is the correlation function of a single site and always stays finite. 
In two dimensions or in the case of a finite lattice, this would lead to a contradiction~\footnote{The sign of $X$ is fixed, so there can be no cancellation of divergences on the right hand side.}, and we indeed know that there should be no phase transition in a finite system or in a two-dimensional system at finite temperature (Mermin-Wagner). 
For $d>2$, the integral
\begin{align}
 \int d\qv' |q'|^{-2} \propto \int_0 d|q'| |q'|^{d-1} |{q}'|^{-2}= \int_0 dq' |q'|^{d-3} \label{eq:integral:dimensionality}
\end{align}
remains finite.
This means that a phase transition can occur while still satisfing Eq.~\eqref{eq:sc:divergence}, but only when taking the limit $N\rightarrow \infty$.

We should point out that many mean-field based methods produce phase transitions in finite systems. In fact, the system size does not even enter traditional Curie-Weiss mean-field theory. The DMFT susceptibility of a finite system can be divergent and non-local extensions like Dual Fermion or Dual Boson do not automatically correct this issue. Here, it is enforced by the self-consistency condition. This situation is in some sense reminiscent of the difference between finite-size and quantum cluster approaches~\cite{Maier05}: finite-size simulations only obtain a phase transition after extrapolating to infinite system size, quantum cluster approaches already find the transition at a finite system size. The former situation is technically correct, the latter is convenient in practical situations.

We recognize that the self-consistency condition serves to enforce the Mermin-Wagner theorem~\cite{Mermin66} in a two-dimensional system, which forbids spontaneous symmetry breaking of continuous symmetries in two-dimensional systems at finite temperature.
For the Hubbard model, the spin rotational symmetry cannot be spontaneously broken.
In particular, this means that there can be no antiferromagnetic state at finite temperature.
On the other hand, at $T=0$, the system is an antiferromagnetic insulator at any $U>0$ if only nearest-neighbor hopping $t_{ij}$ is considered.
Together, this means that at low temperature, the system features very strong and long-ranged antiferromagnetic correlations that are \emph{almost} truly long-ranged ordered. A similar situation takes place in low-dimensional Heisenberg model, see Refs.~\onlinecite{IKK, PhysRevB.39.2344}, and references therein.

This situation, with long but not infinitely ranged correlations, is challenging to reproduces in (computational) approximations.
According to Vilk and Tremblay~\cite{Vilk97}, a sufficient criterion for ensuring that an approximation satisfies the Mermin-Wagner theorem is to verify that the double occupancy $D=\av{n_{\up} n_{\dn}}$ obtained from taking the local, equal-time part of the susceptibility stays within the physical range $[0, n^2/2]$.
The self-consistency condition of the current method ensures that the double occupany of the Hubbard model is equal to that of a reference impurity model that is solved exactly.
The latter stays within the physical bounds, so that the former does as well and the method satisfies the Mermin-Wagner theorem. 

\subsection{Exact properties and approximate solutions}

Let us take a step further back to put these developments in the general context of consistency in approximate solutions to many-body problems and in particular to Hubbard-like models.
The central point is that (many) exact statements can be made about the true, exact solution of the model, to name just a few: equivalence of thermodynamic quantities and response functions according to the Kubo formula; the Mermin-Wagner theorem; conserved quantities corresponding to symmetries and Goldstone modes arising when these symmetries are broken; sum rules and high-frequency asymptotics derived from commutation relations.
\footnote{It is worth observing that these relations occur on very different length and time scales: The Mermin-Wagner theorem and Goldstone modes are long wavelength, low frequency phenomena whereas the high-frequency asymptote of the self-energy is a local, high frequency phenomenon.}
These concepts are central to our understanding of condensed matter physics.

However, approximate solutions are not guaranteed to satisfy these exact properties.
In fact, they will usually not satisfy all these constraints.
This issue goes back to the seminal work of Baym and Kadanoff~\cite{Baym61,Baym62}, who formulated functional constructions to ensure that diagrammatic approaches satisfy certain conservation laws.
The correspondence between correlation and response functions played an important role in the development of the theories of the electron gas~\cite{Geldart70} and of magnetism in itinerant electron systems~\cite{Moriya1973,Dzyaloshinskii1976,Moriya1985}.

In the context of (extensions of) DMFT, the subject of exact properties came up for the thermodynamic consistency of zero- and one-particle properties~\cite{Aichhorn06,Potthoff06} and the consistency between one- and two-particle quantities, in particular the Kubo relation between correlation and response~\cite{Fishman05,Otsuki13b,vanLoon15} and the high-frequency asymptote of the susceptibility~\cite{Krien17}.
Simultaneously, charge conservation and Goldstone modes were investigated~\cite{Hafermann14b,vanLoon14b,Stepanov16,Krien17,Geffroy18}. 

The conclusion of these investigations is that the DMFT approximate solution to the finite-dimensional Hubbard model satisfies \emph{many} of the exact properties, but not \emph{all}.
More specifically, the charge response is consistent~\cite{vanLoon15}, the DMFT susceptibility satisfies global charge conservation~\cite{Hafermann14b,vanLoon14b} and the lowest order terms in the high-frequency asymptotes of the Green's function, (local) self-energy and the (momentum-resolved) susceptibility are consistent with exact relations.
The DMFT functional is conserving~\cite{Krien17} in the sense of Baym and Kadanoff.
On the other hand, the DMFT susceptibility violates the Mermin-Wagner theorem~\footnote{See, e.g., the review~\onlinecite{Rohringer18}.}, the potential energy/double occupancy is inconsistent between the one- and two-particle level\cite{vanLoon16,Rohringer16,Krien17} and
exact relations for the moments of the lower Hubbard band spectral weigh function in the atomic limit are violated~\cite{Esterling18,Esterling18b}.
As we have already seen, DMFT also predicts phase transitions in finite systems, which is inconsistent with fundamental thermodynamic considerations.

Clearly, even the elegant construction of DMFT is not sufficient to recover all exact relations and it is unlikely that any approximate method can.
It is possible to \emph{enforce} specific relations, although usually at a cost.
The self-consistency condition employed here can be seen in this way: it enforces the Mermin-Wagner theorem. In this way, it is similar to the $\lambda$ introduced by Moriya~\cite{Moriya1973,Moriya1985}.

A similar Moriya-$\lambda$ correction plays a central role in ladder-D$\Gamma$A~\cite{Toschi07,Katanin2009}. 
The $\lambda$ of D$\Gamma$A is partially related~\cite{Rohringer18} to the $\Lambda_\omega$ of DB.
Both enter the (inverse) susceptibility as in Eq.~\ref{eq:X}, but in DB $\Lambda$ also enters into the impurity model, as follows from the exact dual transformation, whereas in D$\Gamma$A $\lambda$ correction is introduced by hand and is not included in the impurity model.
We should point out that the instantaneous $\Lambda$ proposed here is even more similar to the instantaneous $\lambda$ of D$\Gamma$A
\cite{Rohringer16}.

\section{Square lattice Hubbard model}

We now turn to numerical investigation of the instantaneous DB.
We study the square lattice Hubbard ($V=0$) model with $t=1$. We use a $32\times 32$ discretization of momentum space. 

\subsection{Effective interaction}

\begin{figure}[b!]
 \includegraphics[width=0.95\linewidth]{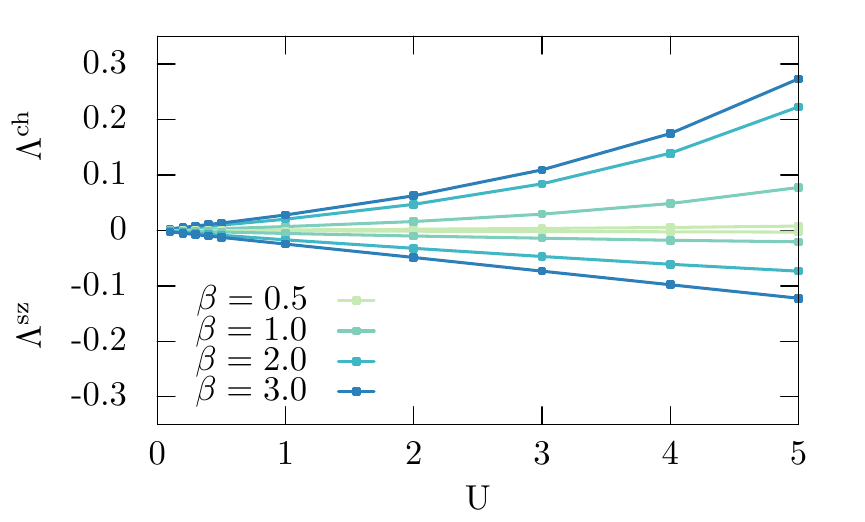}\\
 ~~~\includegraphics[width=0.9\linewidth]{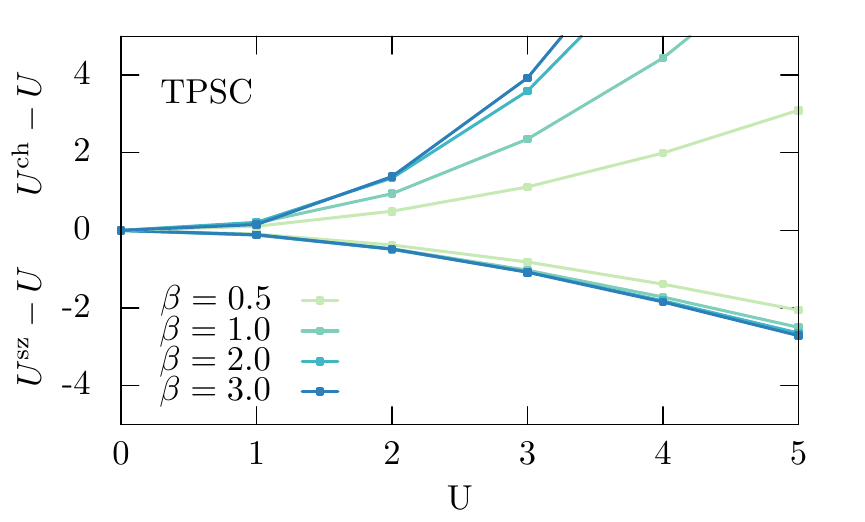}
 \caption{Top panel: Effective interaction as a function of temperature and interaction strength. In all cases shown here, $\Lambda^{\text{ch}}>0$ and $\Lambda^{\text{sz}}<0$. Bottom panel: Effective interactions in the two-particle self-consistent method.}
 \label{fig:effectiveinteraction}
\end{figure}

We start by investigating the effective interactions $\Lambda$, since these are the quantities that enter the impurity model. 
Figure~\ref{fig:effectiveinteraction} shows $\Lambda$ in the small to moderate coupling regime.
We observe that $\Lambda$ depends strongly on the inverse temperature $\beta=1/T$, that $\Lambda^\ch$ and $\Lambda^\sz$ have opposite sign and that $\Lambda$ is proportional to $U$ at small interaction strengths.

The limit of small $U$ can be understood in terms of perturbation theory. 
For the calculation of the susceptibility at small $U$, we can neglect the self-energy and treat the interaction in the RPA fashion, as a geometric series.
Appendix~\ref{app:smallU} gives the details of this approach, which confirm the opposite sign and the proportionality with $U$.
For comparison, the bottom panel of Fig.~\ref{fig:effectiveinteraction} shows the renormalization of the interaction in the two-particle self-consistent approach (TPSC). These calculations have been performed using the TRIQS package~\cite{triqs}.
We observe that the interaction renormalization in Dual Boson is substantially smaller than that in TPSC. The self-consistency in the present approach starts from the DMFT susceptibility, which already includes local self-energy insertions and dynamical vertex corrections. At high temperatures, it provides a good starting point so that almost no renormalization of the impurity interaction is needed ($\Lambda\approx 0$). As the temperature is lowered, the magnitude of $\Lambda$ increases. At the parameters studied here, the magnitude of $\Lambda$ is an order of magnitude smaller than $U^\text{ch/sz}-U$ in TPSC.

\begin{figure}
 \includegraphics{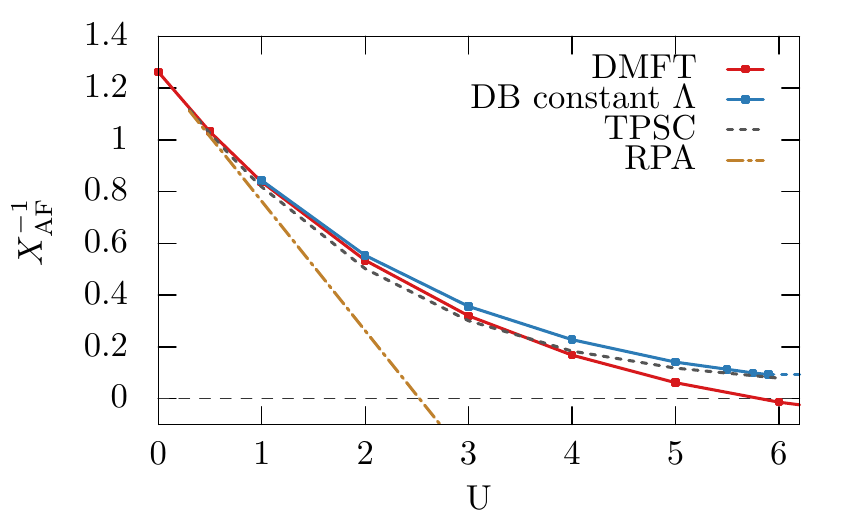}
 \caption{Inverse antiferromagnetic susceptibility $X^\text{AF}_{\qv=(\pi,\pi),\omega=0}$ in the square lattice Hubbard model at $\beta t =3$. A phase transition to an antiferromagnetically ordered phase occurs when this inverse susceptibility is equal to zero (dashed black line). }
 \label{fig:hub:AF}
\end{figure}

One of the main points of the self-consistent Dual Boson approach is that the self-consistency condition enforces the Mermin-Wagner theorem in two dimensions.
The absence of antiferromagnetism in TPSC and the self-consistent Dual Boson approaches is visualized in Fig.~\ref{fig:hub:AF}. 
DMFT is unstable towards antiferromagnetism after $U=6$ when the inverse of the magnetic susceptibility changes sign. The magnetic susceptibility of RPA diverges already at $U\simeq2.5$. The self-consistency condition enforces a positive value for the DB results and thus pushes the inverse susceptibility away from zero. However, enforcing this condition is not easy, since close to an instability the system is very sensitive to small changes in the effective impurity interaction. This makes calculations at $U\geq 6$ extremely unstable.

We proceed with an in-depth look at $U/t=5$, $\beta t=3$, this is just before the antiferromagnetic susceptibility in DMFT diverges. We find $U'/t \approx 5.54$, a more than 10\% increase of the effective interaction. This change in interaction is composed of $\Lambda^\text{ch} \approx 0.41$, $\Lambda^\text{sz}\approx -0.13$.

\begin{figure}[t!]
 \includegraphics{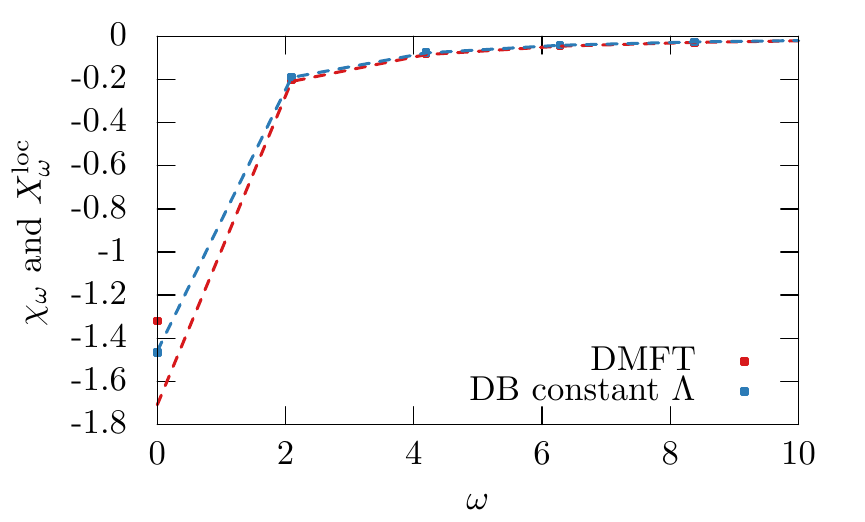}
 \caption{Dynamic susceptibilities at $U=5$ and $\beta=3$, for the square lattice. The points correspond to $\chi$, the lines to $X^\text{loc}$, both in the magnetic channel. 
  }
  \label{fig:2d:chiimp}
\end{figure}

\begin{figure}[h!]
 \includegraphics{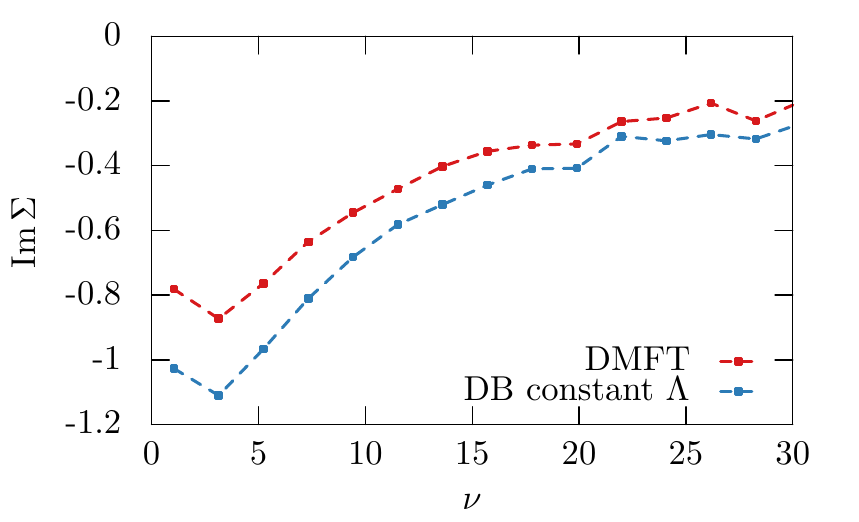}
 \caption{Self-energy at $U=5$ and $\beta=3$, for the square lattice. The points correspond to the Matsubara frequencies $\nu_n$, lines are guides to the eye.
  }
  \label{fig:2d:sigma}
\end{figure}

Looking at the local magnetic susceptibility shown in Fig.~\ref{fig:2d:chiimp}, we see a large inconsistency in DMFT between the impurity and local lattice susceptibility. In the self-consistent approach, this inconsistency is removed almost completely, even though the self-consistency only enforces equality between the frequency-averaged susceptibilities. 
Essentially, the problem in DMFT is that $\abs( X^\text{sz})$ is too large, so that the double occupancy from X can even turn negative. The self-consistency condition solves this by reducing $\abs( X^\text{sz})$. This reduction in the lattice quantity occurs even though $\abs( \chi^\text{sz})$ is increased due to the larger effective impurity interaction.
The larger effective impurity interaction also increases the self-energy, shown in Fig.~\ref{fig:2d:sigma} (see also Appendix~\ref{app:selfenergy}).

\section{Cubic lattice Hubbard model}

We now move to a three-dimensional system and consider $U/t=4$ and $\beta t = 2.5$. We find $U'/t \approx 4.18$, a roughly 5\% change in the effective interaction. The channel decomposition of the effective interaction is $\Lambda^\text{ch}\approx +0.12$ $\Lambda^\text{sz} \approx -0.06$. The difference in magnitude between the charge and spin renormalization shows that we have clearly left the weakly correlated regime. 

The calculations shown here have been performed on a $10\times 10\times 10$ cubic lattice. We have verified that using a $20 \times 20 \times 20$ lattice leads to very similar results. This conforms to the observation that replacing the integral in Eq.~\eqref{eq:integral:dimensionality} by a finite momentum average is a well-behaved operation. All other potential sources of finite size effects are similar to DMFT and not relevant in this parameter regime.

\begin{figure}[t!]
 \includegraphics{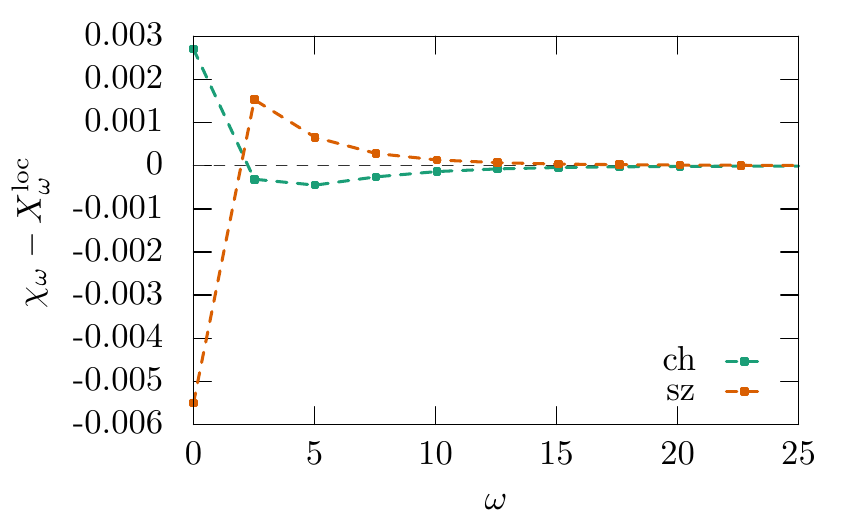}
 \caption{Analysis of the self-consistency condition at $U=4$ and $\beta=2.5$, for a cubic lattice. 
  }
  \label{fig:3d:chiimp}
\end{figure}

\begin{figure}[t!]
 \includegraphics{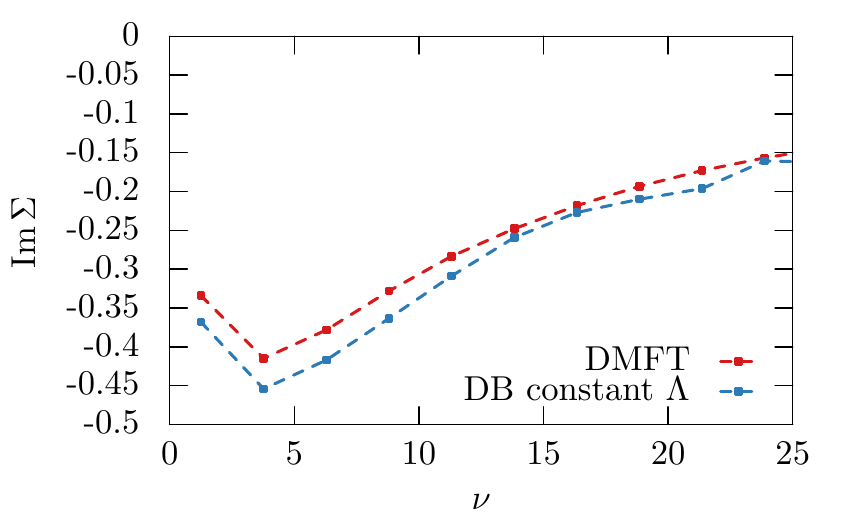}
 \caption{Self-energy at $U=4$ and $\beta=2.5$, for a cubic lattice. The points correspond to the Matsubara frequencies $\nu_n$, lines are guides to the eye.
  }
  \label{fig:3d:sigma}
\end{figure}

The self-consistency condition enforces equality between the frequency averages of $X^\text{loc}$ and $\chi$. In Fig.~\ref{fig:3d:chiimp}, we show the difference of these two susceptibilities as a function of frequency~\footnote{Note that only positive frequencies are shown, so that all finite frequencies have a negative frequency counterpart that is equal by symmetry.} There is an essential difference between finite and zero frequency, in both the charge and the spin channel. The contribution at zero frequency is compensated by the finite frequencies. The sign difference between the charge and spin channel in Fig.~\ref{fig:3d:chiimp} corresponds to the sign difference between $\Lambda^\text{ch}$ and $\Lambda^\text{sz}$.

The self-consistency condition on the two-particle level feeds back to the single-particle level via the impurity model. In Fig.~\ref{fig:3d:sigma} we show the self-energy of the impurity model. We see an enhancement of the self-energy in the self-consistent DB approach. This enhancement originates in the larger effective interaction $U'>U$.

\begin{figure}
 \includegraphics{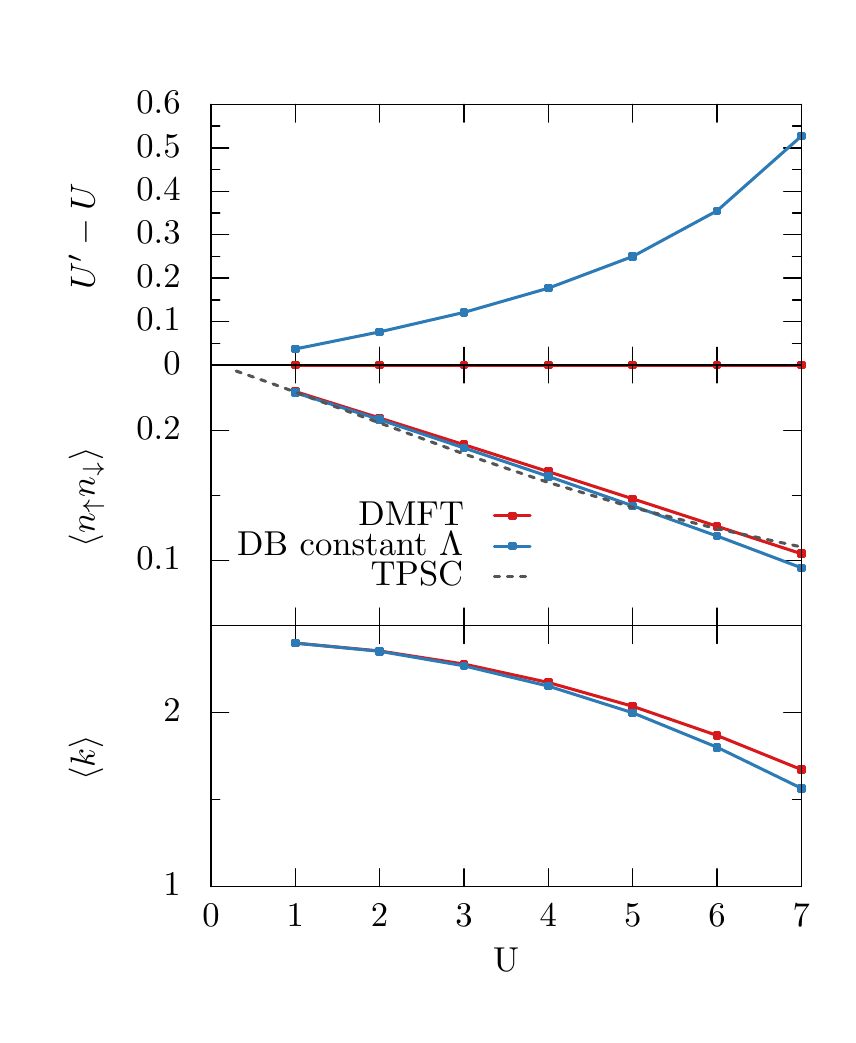}
 \caption{Changes in the impurity model obtained for the cubic lattice at $\beta=2.5$. Effective interaction $U'$, double occupancy $\av{n_\up n_\dn}$ of the impurity model and average perturbation order $\av{k}$ of the CT-HYB solver. The latter two are related to the energy as $E_\text{pot} = U \av{n_\up n_\dn}$ and $E_\text{kin} = - \frac{2}{\beta} \av{k}$~\cite{Haule07}. TPSC result for the double occupancy is shown by a black dashed line.}
  \label{fig:3d:energies}
\end{figure}

Moving away from fixed $U$, Figure~\ref{fig:3d:energies} shows how the effective interaction and the potential and kinetic energy develop as a function of the interaction $U$. The effective interaction becomes more relevant at larger $U$, leading to a reduction in the potential energy and an increase in the kinetic energy~\cite{Haule07} (note the minus sign in $E_\text{kin} = -2 \av{k}/\beta$). We also find that the value of the double occupancy obtained within DMFT and DB theory with constant bosonic hybridization is similar to the one of the TPSC approach.

\section{Extended Hubbard model}

\begin{figure}
 \includegraphics{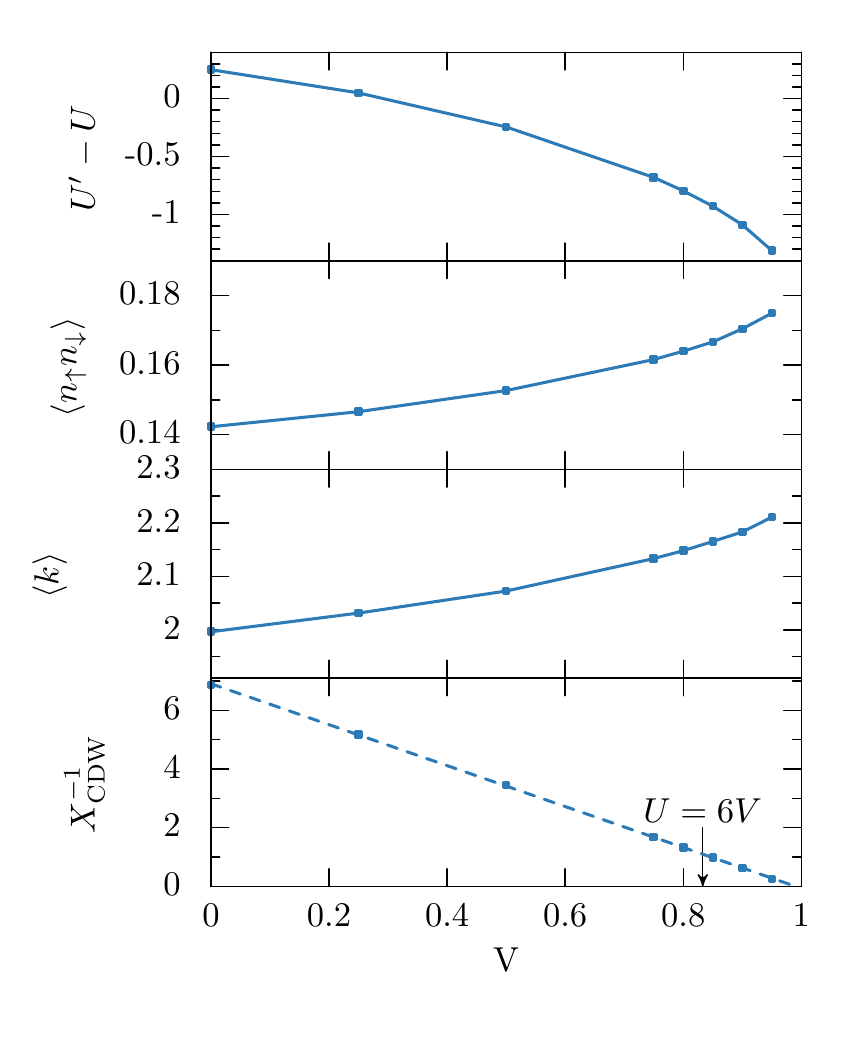}
 \caption{The extended Hubbard model on the cubic lattice, for $U=5$ and $\beta=2.5$. Solid lines are guides to the eye, the dashed line in the bottom panel is a linear fit with intercept at $V \approx 0.99$. The same quantities at $V=0$ are shown in Fig.~\ref{fig:3d:energies}. The bottom panel shows the inverse of the charge-density wave susceptibility. Where this quantity reaches zero, the system becomes unstable towards checkerboard charge order.
  }
  \label{fig:3d:energiesV}
\end{figure}

We now turn our attention to the extended Hubbard model with finite $V$. We study this model on a cubic lattice with fixed local interaction $U=5$ and fixed temperature $\beta=2.5$. We use a $10\times 10 \times 10$ lattice. Figure~\ref{fig:3d:energiesV} shows that screening by the nonlocal interaction $V$ reduces the effective impurity interaction $U'$, so that $U'<U$. As expected, this increases the potential energy and decreases the kinetic energy. Even though it decreases, the impurity interaction is still repulsive for all cases shown here, $U'>0$. Looking at the inverse susceptibility in the charge channel, we find that it approaches zero as $V$ is increased.
Linear extrapolation~\footnote{Note that we find a slope of roughly $X^{-1} \propto -7V$. Without any correlation effects, a slope of $-6V=-V_{q=(\pi,\pi,\pi)}$ would be expected from Eq.~\eqref{eq:X}.} predicts the charge order transition to occur at $V \approx 0.99$.
The arrow marks $V = U/z=0.83$, the point where charge-order becomes favorable in terms of the potential energy~\cite{Vonsovsky_1979}. The actual transition occurs later due to the interplay between potential and kinetic energy and entropy.

The extrapolation here is based on data up to $V=0.95$. In principle, no true divergence is expected in a finite system. This has nothing to do with the Mermin-Wagner theorem, since the charge ordering is not associated with the breaking of a continuous symmetry. Instead, however, for finite-size lattices the self-consistency cycle becomes difficult to stabilize when the phase transition region is approached: small changes in the effective interaction $U'$ will lead to big changes in the susceptibility $X$. 

\begin{figure}
 \includegraphics{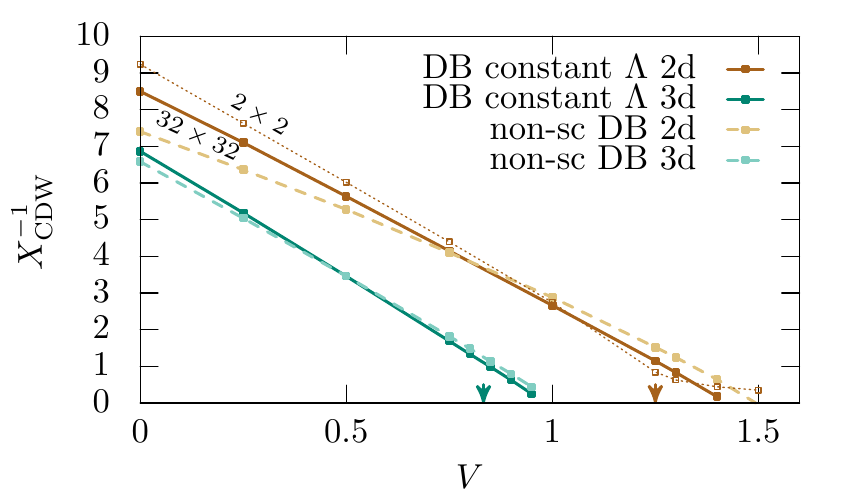}
 \caption{Comparison of the charge order transition in the square and cubic lattice. The square lattice simulations correspond to $\beta=3$, the cubic lattice with $\beta=2.5$, all simulations are at $U=5$. The curves labeled DB constant $\Lambda$ correspond to the method proposed in this work, reference results using the scheme of Ref.~\onlinecite{vanLoon14} are shown as non-sc DB. Arrows indicate $V=U/6$ and $V=U/4$.
  }
  \label{fig:chargeorder}
\end{figure}

We can compare our results obtained with the self-consistent instantaneous interaction with those of Ref.~\cite{vanLoon14}, where a frequency dependent interaction was used and self-consistency was only done at the EDMFT level. Figure~\ref{fig:chargeorder} shows the charge susceptibility for these two methods in the cubic and square lattice Hubbard model. The qualitative behavior of both methods is the same, quantitative differences are visible. The self-consistent approach approaches the phase transition already at smaller values of $U$, that is, the inverse susceptibility is smaller close to the phase transition. The origin of this seems to be the enhanced charge susceptibility of the impurity model that originates in the reduced value of the effective interaction $U'<U$, as was visible in Fig.~\ref{fig:3d:energiesV}.
For the two-dimensional situation, the inverse susceptibility in the self-consistent DB solution is approximately linear in the regime accessible here. Based on Eq.~\eqref{eq:sc:divergence}, the self-consistency condition comes into play only when $X^{-1} \approx 1/L^2 \approx 10^{-3}$ for the $L=32$ system studied here. For a much smaller $L=2$ system (dotted line), it sets in earlier and the dotted line bends upwards at small $V$. Therefore, the true CDW phase boundary can be obtained by extrapolating the results for the inverse charge susceptibility for different system sizes. 

\section{Conclusions and discussion}

We have presented the Dual Boson approach with instantaneous interaction. By construction, this approach produces a susceptibility that satisfies the charge and spin conservation requirement~\cite{Hafermann14b,vanLoon14b,Stepanov16,Krien17,Stepanov18}, and the Mermin-Wagner theorem. The instantaneous interaction assumption means that the method does not need an impurity solver that can handle retarded interactions, an important simplification that makes it more amendable to the simulation of multiband systems. We have illustrated the method in two- and three- dimensional systems and have analyzed how the proposed self-consistency condition affects the appearance of ordered phases. In our method the instantaneous interaction of the impurity model is adjusted by the ``lattice'' self-consistency condition on the bosonic hybridization function. Surprisingly, this constant hybridization function almost completely washes out the difference between the local part of the frequency dependent lattice and impurity susceptibilities. Mean-field based approaches typically overestimate the tendency towards ordered phases, as is clearest in two-dimensional or finite systems where phase transitions are forbidden. Enforcing self-consistency on the two-particle level can cure this deficiency~\cite{Vilk97}. Compared to DMFT (in the absence of the nonlocal interaction $V_{\qv}=0$), we find that the feedback of collective excitations leads to an enhanced effective interaction and to a more correlated impurity model, as is visible in the double occupancy, kinetic energy and self-energy of both two and three-dimensional systems. 

\begin{acknowledgments}
The authors acknowledge useful discussion with Friedrich Krien. 
L.P., E.G.C.P.v.L., M.I.K. and E.A.S. acknowledge support from ERC Advanced Grant 338957 FEMTO/NANO. 
M.I.K. and E.A.S. acknowledge support from the Stichting voor Fundamenteel Onderzoek der Materie (FOM), which is financially supported by the Nederlandse Organisatie voor Wetenschappelijk Onderzoek (NWO).
A.I.L. acknowledges support from the excellence cluster ``The Hamburg Centre for Ultrafast Imaging - Structure, Dynamics and Control of Matter at the Atomic Scale'' and North-German Supercomputing Alliance (HLRN) under the Project No. hhp00042. 
\end{acknowledgments}

\appendix

\section{Analytical results}
\label{app:analytics}

\subsection{$U=0$, $t=0$ or $d=\infty$}

In these three cases the more restrictive self-consistency condition~\cite{Stepanov16} $\chi_\omega = X_{\text{loc}, \omega}$ is fulfilled for all $\omega$ by the solution $\Lambda=0$. 
This means that $\Lambda=0$ is also satisfies the instantaneous self-consistency condition used in this work.
This solution corresponds to the exact DMFT solution of the Hubbard model in these three cases.

\subsection{small $U$}
\label{app:smallU}

We consider the Hubbard model, so $V=0$, and we look at the charge channel of the susceptibility. 
To lowest order in $U$, we can neglect self-energy corrections to the susceptibility and use geometric (RPA-like) equations to describe the impact of the interaction.
The impurity model only knows about the impurity interaction $U'$ and not about $U$ and $\Lambda$ separately,
\begin{align}
 \chi^{-1}_\omega = \chi(U=0)_\omega^{-1} \mp U',
\end{align}
where the sign is $-$ for the charge channel and $+$ for the spin channel.
The dual polarization operator $\tilde{\Pi}$ in the ladder approach is completely determined by the expectation values of the impurity model and by $t$ and $\Delta$, since these enter $\tilde{G}$. In particular, $\tilde{\Pi}$ only depends on $U'$ and not on $U$ and $\Lambda$ separately, and the same holds for $\mathcal{X}$, which is also given by a geometric expression (It is essentially equal to the DMFT susceptibility of a Hubbard model with interaction $U'$.). 
\begin{align}
 \mathcal{X}_{\qv\omega} = \mathcal{X}(U=0)_{\qv\omega}^{-1}\mp U',
\end{align}
The lattice susceptibility according to DB is then given by Eq.~\eqref{eq:X}.
\begin{align}
 X_{\qv\omega}^{-1} = \mathcal{X}_{\qv\omega}^{-1} + \Lambda,
\end{align}
where $\Lambda$ is the interaction in the same channel as the susceptibility.
Combining these equations, for the charge susceptibility we find
\begin{align}
 \left(X^{\text{ch}}_{\qv\omega}\right)^{-1} =& \mathcal{X}(U=0)_{\qv\omega}^{-1} - U'+\Lambda^{\text{ch}} \\
 =& X(U=0)_{\qv\omega}^{-1} - U + \Lambda^{\text{sz}}
\end{align}
where we have used that $\mathcal{X}=X$ and is independent of the channel at $U=0$.
Similarly, we find
\begin{align}
 \left(X^{\text{sz}}_{\qv\omega}\right)^{-1} =& X(U=0)_{\qv\omega}^{-1} + U + \Lambda^{\text{ch}}
\end{align}

Thus, the self-consistency conditions at small $U$ read
\begin{widetext}
\begin{align}
\sum_\omega \frac{\chi(U=0)_\omega}{1-(U+\Lambda^\text{ch}-\Lambda^\text{sz})\chi(U=0)_\omega} = 
\frac{1}{N}\sum_{\qv,\omega} \frac{X(U=0)_{\qv\omega}}{1-(U-\Lambda^\text{sz})X(U=0)_{\qv\omega}}, \\
\sum_\omega \frac{\chi(U=0)_\omega}{1+(U+\Lambda^\text{ch}-\Lambda^\text{sz})\chi(U=0)_\omega} =
\frac{1}{N}\sum_{\qv,\omega} \frac{X(U=0)_{\qv\omega}}{1+(U+\Lambda^\text{ch})X(U=0)_{\qv\omega}}.
\end{align}
\end{widetext}
These two equations can be solved numerically for $\Lambda^\text{ch}$, $\Lambda^\text{sz}$.
Expanding the denominators and using the condition $\chi(U=0)=1/N \sum_\qv X(U=0)$, we find
\begin{align}
& U' \sum_\omega \chi^2(U=0)_\omega = \frac{ U' \pm \Lambda^{\ch/\sz}}{N} \sum_{\qv\omega} X^2(U=0)_{\qv,\omega},\notag \\
 &\Lambda^{\ch/\sz} = \pm U' \cdot C = \pm U \frac{1}{2-C^{-1}}, \label{eq:smallU:lambda}
\end{align} 
where
\begin{align}
 C = \frac{\sum_\omega \chi^2(U=0)_\omega}{\frac{1}{N}\sum_{\qv\omega} X^2(U=0)_{\qv,\omega}}-1.
\end{align}
It is clear that $\Lambda^\ch = - \Lambda^\sz$ in the small $U$ regime described by these equations. 
In addition, both components of $\Lambda$ depend linearly on $U$.
The magnitude of $\Lambda$ depends on $C$. 

\begin{figure}
 \includegraphics{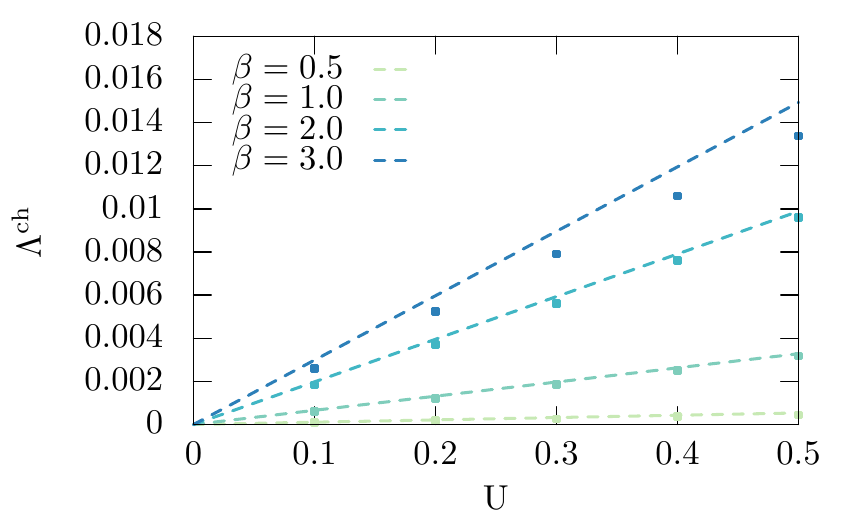}
 \caption{Effective interaction as a function of temperature and interaction strength, with the perturbative result of Eq.~\eqref{eq:smallU:lambda} as the dashed lines and the DB results of Fig.~\ref{fig:effectiveinteraction} as the symbols.}
 \label{fig:effectiveinteractionSMALL}
\end{figure}

The temperature is implicitly contained in $\chi(U=0)$ and $X(U=0)$, which are Lindhardt bubble expressions, and in the sum over Matsubara frequencies. Figure~\ref{fig:effectiveinteractionSMALL} compares the perturbative formula with the numerical results of Fig.~\ref{fig:effectiveinteraction}.

Particle-hole symmetry on hypercubic lattices implies that the change $U\rightarrow -U$ corresponds simply to the interchange of magnetic ($S^{z}$) and density fluctuations, without any difference on the one-particle level. Under the interchange of $\Lambda^\text{ch}$ and $\Lambda^{\text{sz}}$, the expressions given here are clearly consistent with this symmetry.

\section{Self-energy asymptotics}
\label{app:selfenergy}

The general concept that enforcing some exact properties will break others is visible in the present instantaneous DB as well.
An example is the high-frequency asymptote of the local self-energy of the paramagnetic Hubbard model, 
\begin{align}
 \Sigma^{\text{lat}}_\nu \overset{\nu\rightarrow\infty}{=} \frac{U\av{n}}{2} +\frac{U^2 \av{n}^2}{4}\frac{1}{i\nu}+\ldots.
\end{align}
For the impurity model, with interaction $U'\neq U$, we find
\begin{align}
 \Sigma^\text{imp}_\nu \overset{\nu\rightarrow\infty}{=} \frac{U' \av{n}}{2} +\frac{U'^2 \av{n}^2}{4}\frac{1}{i\nu}+\ldots.
\end{align}
At high frequencies, where the denominator $1+g\tilde{\Sigma}$ is equal to unity, the relation between dual and lattice self-energy is $\Sigma^\text{lat} = \Sigma^\text{imp} + \tilde{\Sigma}$.
This means that the \emph{exact} solution of the dual action should have
\begin{align}
 \tilde{\Sigma}^\text{exact}_\nu \overset{\nu\rightarrow\infty}{=} \frac{(U-U') \av{n}}{2} +\frac{(U^2-U'^2) \av{n}^2 }{4} \frac{1}{i\nu}+\ldots.
\end{align}
Let us see how this exact expression can arise in dual perturbation theory. We consider a situation where the difference $U'-U=\Lambda$ is small, so that dual perturbation theory is justified and we also consider $U'$ sufficiently small that the vertices can be simplified by doing perturbation theory in the impurity model.

The bare bosonic dual propagator simplifies in the Hubbard model with instantaneous impurity interaction, it is local~\footnote{For the Hubbard model, $\tilde{X}^{(0)}$ is always local. The self-consistency condition $\sum_\qv \tilde{X}^{(0)}=0$ then automatically ensures that $\tilde{X}=0$ everywhere and that Dual Boson reduces to Dual Fermion. The self-consistency condition used in this work does not require $\sum_\qv \tilde{X}^{(0)}=0$ and $\tilde{X}$ is finite in the Hubbard model.} and given by,
\begin{align}
 \tilde{X}^{(0)}_{\qv\omega} &= \left( \chi^{-1}_\omega + \Lambda \right)^{-1} -\chi_\omega \\
 \tilde{X}^{(0)}_{\qv\omega}/\chi_\omega^2 &= \frac{\Lambda}{1+\Lambda \chi_\omega} \\
 &\approx \Lambda=U'-U. \label{eq:app:xdual}
\end{align}
Here, the last line is obtained since we are interested in small $\Lambda$. This shows that the number of bosonic propagators determines the order in $(U'-U)$.

\begin{figure}
\begin{tikzpicture}

 \node at (-1,0.5) {$\tilde{\Sigma}_\nu=$};

 \draw[thick,fill=gray] (0,0) -- (1,0) -- (1,1) -- (0,1) -- cycle ;

 \node at (0.5,0.5) {$\gamma^{(2,2)}_{\nu\nu\omega\omega}$} ;

 \node[circle,draw=black,fill=red] (northwestern) at (1,1) {} ;
 \node[circle,draw=black,fill=red] (northeastern) at (0,1) {} ;

 \draw[draw,thick,dashed,out=90,in=90] (northwestern) to node[above] {$\tilde{X}_\omega$} (northeastern) ;

 \draw[thick,->] (-0.5,0) -- (0,0);
 \draw[thick,->] (1,0) -- (1.5,0);

\end{tikzpicture}
\,\,
\begin{tikzpicture}
 \node[anchor=east] at (-0.5,0.5) {$\gamma^{(2,2)}_{\nu_1 \nu_2\omega_1\omega_2} \sim$} ;

 \node[circle,draw=black,fill=red] (northeastern) at (1,1) {} ;
 \node[circle,draw=black,fill=red] (northwestern) at (0,1) {} ;

 \draw[->] (0,0) .. controls +(0.4,0.4) and +(0.4,-0.4) .. node[left] {$g_{\nu_1}$} (northwestern) ;
 \draw[<-] (1,0) .. controls +(-0.4,0.4) and +(-0.4,-0.4) .. node[right] {$g_{\nu_2}$} (northeastern) ;
 \draw[->] (northwestern) -- node[above] {$g_{\nu_3}$} (northeastern) ;
\end{tikzpicture}
\caption{Left: Self-energy diagram that contributes to the asymptote. Right: Diagram of trivial contribution to $\gamma^{2,2}$.}
\label{fig:app:selfenergy}
\end{figure}
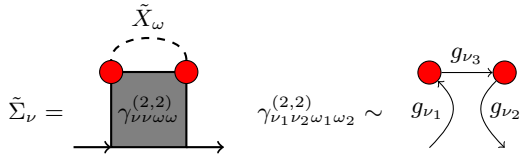

We identify three basic diagrams that could be relevant for the asymptote: a ``Hartree'' and a ``Fock'' diagram both containing two two-fermion-one-boson vertices and a diagram with a single two-fermion-two-boson vertex $\gamma^{(2,2)}$.
The first two of these diagrams vanish since they contain a local Dual Fermion propagator, which is zero by the fermionic self-consistency condition. It is the third diagram, shown in Fig.~\ref{fig:app:selfenergy}, that we are interested in here. To evaluate it, we need to find a simplified expression for the vertex.
In Dual Boson, vertices involving the bosonic degree of freedom typically have a ``trivial'' contribution, as has been discussed at length for the fermion-boson vertex~\cite{Rubtsov12,vanLoon14,vanLoon18}. It essentially originates in the fact that the number of Wick contractions is reduced when going from $c^\dagger c$ to $n$. 
This also applies to $\gamma^{(2,2)}$, which has a trivial contribution
\begin{align}
  \gamma^{(2,2)}_{\nu_1 \nu_2 \omega_1 \omega_2} \times \chi_{\omega_1}\chi_{\omega_2} \sim g_{\nu_3} \, \delta_{\omega_1 \omega_2} \delta_{\nu_1 \nu_2} \delta_{\nu_3-\nu_1+\omega_1}. \label{eq:vertex6}
\end{align}
Evaluating the self-energy with symmetry factor $\frac{1}{2}$, and combining Eqs.~\eqref{eq:app:xdual} and \eqref{eq:vertex6} gives
\begin{align}
 \tilde{\Sigma}_{\nu} 
      =& \frac{1}{2} \sum \tilde{X_{\omega}} \cdot \gamma^{(2,2)}_{\nu \nu \omega \omega} \\
      \approx& \frac{1}{2} \sum_{\sigma\nu} (U'-U) g_{\sigma\nu} \notag\\
      =& (U'-U) \frac{\av{n}}{2}, \notag
\end{align}
exactly the desired expression for the first order term.

Higher-order terms should appear in a similar fashion, although they become progressively harder to evaluate: multiple diagrams will contribute and the diagrams become more involved. 

The interpretation of this result is two-fold. On the one hand, as long as $U'-U$ is not too large, deviations between the exact and numerical asymptotics of the self-energy will also stay small. On the other hand, it shows that a diagram that includes a three-particle vertex is already needed to recover only the lowest order term in the asymptotic expansion of the self-energy. This occurence of three-particle vertices to achieve consistency is reminiscent of Refs.~\cite{vanLoon15,Krien17,vanLoon18b}.

\bibliography{references}

\begin{thebibliography}{73}%
\makeatletter
\providecommand \@ifxundefined [1]{%
 \@ifx{#1\undefined}
}%
\providecommand \@ifnum [1]{%
 \ifnum #1\expandafter \@firstoftwo
 \else \expandafter \@secondoftwo
 \fi
}%
\providecommand \@ifx [1]{%
 \ifx #1\expandafter \@firstoftwo
 \else \expandafter \@secondoftwo
 \fi
}%
\providecommand \natexlab [1]{#1}%
\providecommand \enquote  [1]{``#1''}%
\providecommand \bibnamefont  [1]{#1}%
\providecommand \bibfnamefont [1]{#1}%
\providecommand \citenamefont [1]{#1}%
\providecommand \href@noop [0]{\@secondoftwo}%
\providecommand \href [0]{\begingroup \@sanitize@url \@href}%
\providecommand \@href[1]{\@@startlink{#1}\@@href}%
\providecommand \@@href[1]{\endgroup#1\@@endlink}%
\providecommand \@sanitize@url [0]{\catcode `\\12\catcode `\$12\catcode
  `\&12\catcode `\#12\catcode `\^12\catcode `\_12\catcode `\%12\relax}%
\providecommand \@@startlink[1]{}%
\providecommand \@@endlink[0]{}%
\providecommand \url  [0]{\begingroup\@sanitize@url \@url }%
\providecommand \@url [1]{\endgroup\@href {#1}{\urlprefix }}%
\providecommand \urlprefix  [0]{URL }%
\providecommand \Eprint [0]{\href }%
\providecommand \doibase [0]{http://dx.doi.org/}%
\providecommand \selectlanguage [0]{\@gobble}%
\providecommand \bibinfo  [0]{\@secondoftwo}%
\providecommand \bibfield  [0]{\@secondoftwo}%
\providecommand \translation [1]{[#1]}%
\providecommand \BibitemOpen [0]{}%
\providecommand \bibitemStop [0]{}%
\providecommand \bibitemNoStop [0]{.\EOS\space}%
\providecommand \EOS [0]{\spacefactor3000\relax}%
\providecommand \BibitemShut  [1]{\csname bibitem#1\endcsname}%
\let\auto@bib@innerbib\@empty
%</preamble>
\bibitem [{\citenamefont {Hubbard}(1963)}]{Hubbard63}%
  \BibitemOpen
  \bibfield  {author} {\bibinfo {author} {\bibfnamefont {J.}~\bibnamefont
  {Hubbard}},\ }\href {\doibase 10.1098/rspa.1963.0204} {\bibfield  {journal}
  {\bibinfo  {journal} {Proc. R. Soc. A.}\ }\textbf {\bibinfo {volume} {276}},\
  \bibinfo {pages} {238} (\bibinfo {year} {1963})}\BibitemShut {NoStop}%
\bibitem [{\citenamefont {Kanamori}(1963)}]{Kanamori63}%
  \BibitemOpen
  \bibfield  {author} {\bibinfo {author} {\bibfnamefont {J.}~\bibnamefont
  {Kanamori}},\ }\href@noop {} {\bibfield  {journal} {\bibinfo  {journal}
  {Prog. Theor. Phys.}\ }\textbf {\bibinfo {volume} {30}},\ \bibinfo {pages}
  {275} (\bibinfo {year} {1963})}\BibitemShut {NoStop}%
\bibitem [{\citenamefont {Gutzwiller}(1963)}]{Gutzwiller63}%
  \BibitemOpen
  \bibfield  {author} {\bibinfo {author} {\bibfnamefont {M.~C.}\ \bibnamefont
  {Gutzwiller}},\ }\href {\doibase 10.1103/PhysRevLett.10.159} {\bibfield
  {journal} {\bibinfo  {journal} {Phys. Rev. Lett.}\ }\textbf {\bibinfo
  {volume} {10}},\ \bibinfo {pages} {159} (\bibinfo {year} {1963})}\BibitemShut
  {NoStop}%
\bibitem [{\citenamefont {Hubbard}(1964)}]{Hubbard64}%
  \BibitemOpen
  \bibfield  {author} {\bibinfo {author} {\bibfnamefont {J.}~\bibnamefont
  {Hubbard}},\ }\href {\doibase 10.1098/rspa.1964.0190} {\bibfield  {journal}
  {\bibinfo  {journal} {Proc. R. Soc. A.}\ }\textbf {\bibinfo {volume} {281}},\
  \bibinfo {pages} {401} (\bibinfo {year} {1964})}\BibitemShut {NoStop}%
\bibitem [{\citenamefont {Georges}\ \emph {et~al.}(1996)\citenamefont
  {Georges}, \citenamefont {Kotliar}, \citenamefont {Krauth},\ and\
  \citenamefont {Rozenberg}}]{Georges96}%
  \BibitemOpen
  \bibfield  {author} {\bibinfo {author} {\bibfnamefont {A.}~\bibnamefont
  {Georges}}, \bibinfo {author} {\bibfnamefont {G.}~\bibnamefont {Kotliar}},
  \bibinfo {author} {\bibfnamefont {W.}~\bibnamefont {Krauth}}, \ and\ \bibinfo
  {author} {\bibfnamefont {M.~J.}\ \bibnamefont {Rozenberg}},\ }\href {\doibase
  10.1103/RevModPhys.68.13} {\bibfield  {journal} {\bibinfo  {journal} {Rev.
  Mod. Phys.}\ }\textbf {\bibinfo {volume} {68}},\ \bibinfo {pages} {13}
  (\bibinfo {year} {1996})}\BibitemShut {NoStop}%
\bibitem [{\citenamefont {Metzner}\ and\ \citenamefont
  {Vollhardt}(1989)}]{Metzner89}%
  \BibitemOpen
  \bibfield  {author} {\bibinfo {author} {\bibfnamefont {W.}~\bibnamefont
  {Metzner}}\ and\ \bibinfo {author} {\bibfnamefont {D.}~\bibnamefont
  {Vollhardt}},\ }\href {\doibase 10.1103/PhysRevLett.62.324} {\bibfield
  {journal} {\bibinfo  {journal} {Phys. Rev. Lett.}\ }\textbf {\bibinfo
  {volume} {62}},\ \bibinfo {pages} {324} (\bibinfo {year} {1989})}\BibitemShut
  {NoStop}%
\bibitem [{\citenamefont {Rohringer}\ \emph {et~al.}(2018)\citenamefont
  {Rohringer}, \citenamefont {Hafermann}, \citenamefont {Toschi}, \citenamefont
  {Katanin}, \citenamefont {Antipov}, \citenamefont {Katsnelson}, \citenamefont
  {Lichtenstein}, \citenamefont {Rubtsov},\ and\ \citenamefont
  {Held}}]{Rohringer18}%
  \BibitemOpen
  \bibfield  {author} {\bibinfo {author} {\bibfnamefont {G.}~\bibnamefont
  {Rohringer}}, \bibinfo {author} {\bibfnamefont {H.}~\bibnamefont
  {Hafermann}}, \bibinfo {author} {\bibfnamefont {A.}~\bibnamefont {Toschi}},
  \bibinfo {author} {\bibfnamefont {A.~A.}\ \bibnamefont {Katanin}}, \bibinfo
  {author} {\bibfnamefont {A.~E.}\ \bibnamefont {Antipov}}, \bibinfo {author}
  {\bibfnamefont {M.~I.}\ \bibnamefont {Katsnelson}}, \bibinfo {author}
  {\bibfnamefont {A.~I.}\ \bibnamefont {Lichtenstein}}, \bibinfo {author}
  {\bibfnamefont {A.~N.}\ \bibnamefont {Rubtsov}}, \ and\ \bibinfo {author}
  {\bibfnamefont {K.}~\bibnamefont {Held}},\ }\href {\doibase
  10.1103/RevModPhys.90.025003} {\bibfield  {journal} {\bibinfo  {journal}
  {Rev. Mod. Phys.}\ }\textbf {\bibinfo {volume} {90}},\ \bibinfo {pages}
  {025003} (\bibinfo {year} {2018})}\BibitemShut {NoStop}%
\bibitem [{\citenamefont {Sengupta}\ and\ \citenamefont
  {Georges}(1995)}]{Sengupta95}%
  \BibitemOpen
  \bibfield  {author} {\bibinfo {author} {\bibfnamefont {A.~M.}\ \bibnamefont
  {Sengupta}}\ and\ \bibinfo {author} {\bibfnamefont {A.}~\bibnamefont
  {Georges}},\ }\href {\doibase 10.1103/PhysRevB.52.10295} {\bibfield
  {journal} {\bibinfo  {journal} {Phys. Rev. B}\ }\textbf {\bibinfo {volume}
  {52}},\ \bibinfo {pages} {10295} (\bibinfo {year} {1995})}\BibitemShut
  {NoStop}%
\bibitem [{\citenamefont {Si}\ and\ \citenamefont {Smith}(1996)}]{Si96}%
  \BibitemOpen
  \bibfield  {author} {\bibinfo {author} {\bibfnamefont {Q.}~\bibnamefont
  {Si}}\ and\ \bibinfo {author} {\bibfnamefont {J.~L.}\ \bibnamefont {Smith}},\
  }\href {\doibase 10.1103/PhysRevLett.77.3391} {\bibfield  {journal} {\bibinfo
   {journal} {Phys. Rev. Lett.}\ }\textbf {\bibinfo {volume} {77}},\ \bibinfo
  {pages} {3391} (\bibinfo {year} {1996})}\BibitemShut {NoStop}%
\bibitem [{\citenamefont {Kajueter}(1996)}]{Kajueter96}%
  \BibitemOpen
  \bibfield  {author} {\bibinfo {author} {\bibfnamefont {H.}~\bibnamefont
  {Kajueter}},\ }\emph {\bibinfo {title} {Interpolating Perturbation Scheme for
  Correlated Electron Systems}},\ \href@noop {} {Ph.D. thesis},\ \bibinfo
  {school} {Rutgers University} (\bibinfo {year} {1996})\BibitemShut {NoStop}%
\bibitem [{\citenamefont {Smith}\ and\ \citenamefont {Si}(2000)}]{Smith00}%
  \BibitemOpen
  \bibfield  {author} {\bibinfo {author} {\bibfnamefont {J.~L.}\ \bibnamefont
  {Smith}}\ and\ \bibinfo {author} {\bibfnamefont {Q.}~\bibnamefont {Si}},\
  }\href {\doibase 10.1103/PhysRevB.61.5184} {\bibfield  {journal} {\bibinfo
  {journal} {Phys. Rev. B}\ }\textbf {\bibinfo {volume} {61}},\ \bibinfo
  {pages} {5184} (\bibinfo {year} {2000})}\BibitemShut {NoStop}%
\bibitem [{\citenamefont {Chitra}\ and\ \citenamefont
  {Kotliar}(2000)}]{Chitra00}%
  \BibitemOpen
  \bibfield  {author} {\bibinfo {author} {\bibfnamefont {R.}~\bibnamefont
  {Chitra}}\ and\ \bibinfo {author} {\bibfnamefont {G.}~\bibnamefont
  {Kotliar}},\ }\href {\doibase 10.1103/PhysRevLett.84.3678} {\bibfield
  {journal} {\bibinfo  {journal} {Phys. Rev. Lett.}\ }\textbf {\bibinfo
  {volume} {84}},\ \bibinfo {pages} {3678} (\bibinfo {year}
  {2000})}\BibitemShut {NoStop}%
\bibitem [{\citenamefont {Chitra}\ and\ \citenamefont
  {Kotliar}(2001)}]{Chitra01}%
  \BibitemOpen
  \bibfield  {author} {\bibinfo {author} {\bibfnamefont {R.}~\bibnamefont
  {Chitra}}\ and\ \bibinfo {author} {\bibfnamefont {G.}~\bibnamefont
  {Kotliar}},\ }\href {\doibase 10.1103/PhysRevB.63.115110} {\bibfield
  {journal} {\bibinfo  {journal} {Phys. Rev. B}\ }\textbf {\bibinfo {volume}
  {63}},\ \bibinfo {pages} {115110} (\bibinfo {year} {2001})}\BibitemShut
  {NoStop}%
\bibitem [{\citenamefont {Rubtsov}\ \emph {et~al.}(2012)\citenamefont
  {Rubtsov}, \citenamefont {Katsnelson},\ and\ \citenamefont
  {Lichtenstein}}]{Rubtsov12}%
  \BibitemOpen
  \bibfield  {author} {\bibinfo {author} {\bibfnamefont {A.~N.}\ \bibnamefont
  {Rubtsov}}, \bibinfo {author} {\bibfnamefont {M.~I.}\ \bibnamefont
  {Katsnelson}}, \ and\ \bibinfo {author} {\bibfnamefont {A.~I.}\ \bibnamefont
  {Lichtenstein}},\ }\href {\doibase 10.1016/j.aop.2012.01.002} {\bibfield
  {journal} {\bibinfo  {journal} {Annals of Physics}\ }\textbf {\bibinfo
  {volume} {327}},\ \bibinfo {pages} {1320} (\bibinfo {year}
  {2012})}\BibitemShut {NoStop}%
\bibitem [{\citenamefont {Stepanov}\ \emph {et~al.}(2016)\citenamefont
  {Stepanov}, \citenamefont {van Loon}, \citenamefont {Katanin}, \citenamefont
  {Lichtenstein}, \citenamefont {Katsnelson},\ and\ \citenamefont
  {Rubtsov}}]{Stepanov16}%
  \BibitemOpen
  \bibfield  {author} {\bibinfo {author} {\bibfnamefont {E.~A.}\ \bibnamefont
  {Stepanov}}, \bibinfo {author} {\bibfnamefont {E.~G. C.~P.}\ \bibnamefont
  {van Loon}}, \bibinfo {author} {\bibfnamefont {A.~A.}\ \bibnamefont
  {Katanin}}, \bibinfo {author} {\bibfnamefont {A.~I.}\ \bibnamefont
  {Lichtenstein}}, \bibinfo {author} {\bibfnamefont {M.~I.}\ \bibnamefont
  {Katsnelson}}, \ and\ \bibinfo {author} {\bibfnamefont {A.~N.}\ \bibnamefont
  {Rubtsov}},\ }\href {\doibase 10.1103/PhysRevB.93.045107} {\bibfield
  {journal} {\bibinfo  {journal} {Phys. Rev. B}\ }\textbf {\bibinfo {volume}
  {93}},\ \bibinfo {pages} {045107} (\bibinfo {year} {2016})}\BibitemShut
  {NoStop}%
\bibitem [{\citenamefont {Otsuki}(2013)}]{Otsuki13}%
  \BibitemOpen
  \bibfield  {author} {\bibinfo {author} {\bibfnamefont {J.}~\bibnamefont
  {Otsuki}},\ }\href {\doibase 10.1103/PhysRevB.87.125102} {\bibfield
  {journal} {\bibinfo  {journal} {Phys. Rev. B}\ }\textbf {\bibinfo {volume}
  {87}},\ \bibinfo {pages} {125102} (\bibinfo {year} {2013})}\BibitemShut
  {NoStop}%
\bibitem [{\citenamefont {Steiner}\ \emph {et~al.}(2015)\citenamefont
  {Steiner}, \citenamefont {Nomura},\ and\ \citenamefont {Werner}}]{Steiner15}%
  \BibitemOpen
  \bibfield  {author} {\bibinfo {author} {\bibfnamefont {K.}~\bibnamefont
  {Steiner}}, \bibinfo {author} {\bibfnamefont {Y.}~\bibnamefont {Nomura}}, \
  and\ \bibinfo {author} {\bibfnamefont {P.}~\bibnamefont {Werner}},\ }\href
  {\doibase 10.1103/PhysRevB.92.115123} {\bibfield  {journal} {\bibinfo
  {journal} {Phys. Rev. B}\ }\textbf {\bibinfo {volume} {92}},\ \bibinfo
  {pages} {115123} (\bibinfo {year} {2015})}\BibitemShut {NoStop}%
\bibitem [{\citenamefont {Werner}\ and\ \citenamefont
  {Millis}(2007)}]{Werner07}%
  \BibitemOpen
  \bibfield  {author} {\bibinfo {author} {\bibfnamefont {P.}~\bibnamefont
  {Werner}}\ and\ \bibinfo {author} {\bibfnamefont {A.~J.}\ \bibnamefont
  {Millis}},\ }\href {\doibase 10.1103/PhysRevLett.99.146404} {\bibfield
  {journal} {\bibinfo  {journal} {Phys. Rev. Lett.}\ }\textbf {\bibinfo
  {volume} {99}},\ \bibinfo {pages} {146404} (\bibinfo {year}
  {2007})}\BibitemShut {NoStop}%
\bibitem [{\citenamefont {Werner}\ and\ \citenamefont
  {Millis}(2010)}]{Werner10}%
  \BibitemOpen
  \bibfield  {author} {\bibinfo {author} {\bibfnamefont {P.}~\bibnamefont
  {Werner}}\ and\ \bibinfo {author} {\bibfnamefont {A.~J.}\ \bibnamefont
  {Millis}},\ }\href {\doibase 10.1103/PhysRevLett.104.146401} {\bibfield
  {journal} {\bibinfo  {journal} {Phys. Rev. Lett.}\ }\textbf {\bibinfo
  {volume} {104}},\ \bibinfo {pages} {146401} (\bibinfo {year}
  {2010})}\BibitemShut {NoStop}%
\bibitem [{\citenamefont {Krien}\ \emph {et~al.}(2017)\citenamefont {Krien},
  \citenamefont {van Loon}, \citenamefont {Hafermann}, \citenamefont {Otsuki},
  \citenamefont {Katsnelson},\ and\ \citenamefont {Lichtenstein}}]{Krien17}%
  \BibitemOpen
  \bibfield  {author} {\bibinfo {author} {\bibfnamefont {F.}~\bibnamefont
  {Krien}}, \bibinfo {author} {\bibfnamefont {E.~G. C.~P.}\ \bibnamefont {van
  Loon}}, \bibinfo {author} {\bibfnamefont {H.}~\bibnamefont {Hafermann}},
  \bibinfo {author} {\bibfnamefont {J.}~\bibnamefont {Otsuki}}, \bibinfo
  {author} {\bibfnamefont {M.~I.}\ \bibnamefont {Katsnelson}}, \ and\ \bibinfo
  {author} {\bibfnamefont {A.~I.}\ \bibnamefont {Lichtenstein}},\ }\href
  {\doibase 10.1103/PhysRevB.96.075155} {\bibfield  {journal} {\bibinfo
  {journal} {Phys. Rev. B}\ }\textbf {\bibinfo {volume} {96}},\ \bibinfo
  {pages} {075155} (\bibinfo {year} {2017})}\BibitemShut {NoStop}%
\bibitem [{\citenamefont {Stepanov}\ \emph {et~al.}(2018)\citenamefont
  {Stepanov}, \citenamefont {Peters}, \citenamefont {Krivenko}, \citenamefont
  {Lichtenstein}, \citenamefont {Katsnelson},\ and\ \citenamefont
  {Rubtsov}}]{Stepanov18}%
  \BibitemOpen
  \bibfield  {author} {\bibinfo {author} {\bibfnamefont {E.~A.}\ \bibnamefont
  {Stepanov}}, \bibinfo {author} {\bibfnamefont {L.}~\bibnamefont {Peters}},
  \bibinfo {author} {\bibfnamefont {I.~S.}\ \bibnamefont {Krivenko}}, \bibinfo
  {author} {\bibfnamefont {A.~I.}\ \bibnamefont {Lichtenstein}}, \bibinfo
  {author} {\bibfnamefont {M.~I.}\ \bibnamefont {Katsnelson}}, \ and\ \bibinfo
  {author} {\bibfnamefont {A.~N.}\ \bibnamefont {Rubtsov}},\ }\href {\doibase
  10.1038/s41535-018-0128-x} {\bibfield  {journal} {\bibinfo  {journal} {npj
  Quantum Materials}\ }\textbf {\bibinfo {volume} {3}},\ \bibinfo {pages} {54}
  (\bibinfo {year} {2018})}\BibitemShut {NoStop}%
\bibitem [{\citenamefont {Vilk}\ \emph {et~al.}(1994)\citenamefont {Vilk},
  \citenamefont {Chen},\ and\ \citenamefont {Tremblay}}]{Vilk94}%
  \BibitemOpen
  \bibfield  {author} {\bibinfo {author} {\bibfnamefont {Y.~M.}\ \bibnamefont
  {Vilk}}, \bibinfo {author} {\bibfnamefont {L.}~\bibnamefont {Chen}}, \ and\
  \bibinfo {author} {\bibfnamefont {A.-M.~S.}\ \bibnamefont {Tremblay}},\
  }\href {\doibase 10.1103/PhysRevB.49.13267} {\bibfield  {journal} {\bibinfo
  {journal} {Phys. Rev. B}\ }\textbf {\bibinfo {volume} {49}},\ \bibinfo
  {pages} {13267} (\bibinfo {year} {1994})}\BibitemShut {NoStop}%
\bibitem [{\citenamefont {{Y.M. Vilk}}\ and\ \citenamefont {{A.-M.S.
  Tremblay}}(1997)}]{Vilk97}%
  \BibitemOpen
  \bibfield  {author} {\bibinfo {author} {\bibnamefont {{Y.M. Vilk}}}\ and\
  \bibinfo {author} {\bibnamefont {{A.-M.S. Tremblay}}},\ }\href {\doibase
  10.1051/jp1:1997135} {\bibfield  {journal} {\bibinfo  {journal} {J. Phys. I
  France}\ }\textbf {\bibinfo {volume} {7}},\ \bibinfo {pages} {1309} (\bibinfo
  {year} {1997})}\BibitemShut {NoStop}%
\bibitem [{\citenamefont {Toschi}\ \emph {et~al.}(2007)\citenamefont {Toschi},
  \citenamefont {Katanin},\ and\ \citenamefont {Held}}]{Toschi07}%
  \BibitemOpen
  \bibfield  {author} {\bibinfo {author} {\bibfnamefont {A.}~\bibnamefont
  {Toschi}}, \bibinfo {author} {\bibfnamefont {A.~A.}\ \bibnamefont {Katanin}},
  \ and\ \bibinfo {author} {\bibfnamefont {K.}~\bibnamefont {Held}},\ }\href
  {\doibase 10.1103/PhysRevB.75.045118} {\bibfield  {journal} {\bibinfo
  {journal} {Phys. Rev. B}\ }\textbf {\bibinfo {volume} {75}},\ \bibinfo
  {pages} {045118} (\bibinfo {year} {2007})}\BibitemShut {NoStop}%
\bibitem [{\citenamefont {Rubtsov}\ \emph {et~al.}(2008)\citenamefont
  {Rubtsov}, \citenamefont {Katsnelson},\ and\ \citenamefont
  {Lichtenstein}}]{Rubtsov08}%
  \BibitemOpen
  \bibfield  {author} {\bibinfo {author} {\bibfnamefont {A.~N.}\ \bibnamefont
  {Rubtsov}}, \bibinfo {author} {\bibfnamefont {M.~I.}\ \bibnamefont
  {Katsnelson}}, \ and\ \bibinfo {author} {\bibfnamefont {A.~I.}\ \bibnamefont
  {Lichtenstein}},\ }\href {\doibase 10.1103/PhysRevB.77.033101} {\bibfield
  {journal} {\bibinfo  {journal} {Phys. Rev. B}\ }\textbf {\bibinfo {volume}
  {77}},\ \bibinfo {pages} {033101} (\bibinfo {year} {2008})}\BibitemShut
  {NoStop}%
\bibitem [{\citenamefont {van Loon}\ \emph
  {et~al.}(2014{\natexlab{a}})\citenamefont {van Loon}, \citenamefont
  {Lichtenstein}, \citenamefont {Katsnelson}, \citenamefont {Parcollet},\ and\
  \citenamefont {Hafermann}}]{vanLoon14}%
  \BibitemOpen
  \bibfield  {author} {\bibinfo {author} {\bibfnamefont {E.~G. C.~P.}\
  \bibnamefont {van Loon}}, \bibinfo {author} {\bibfnamefont {A.~I.}\
  \bibnamefont {Lichtenstein}}, \bibinfo {author} {\bibfnamefont {M.~I.}\
  \bibnamefont {Katsnelson}}, \bibinfo {author} {\bibfnamefont
  {O.}~\bibnamefont {Parcollet}}, \ and\ \bibinfo {author} {\bibfnamefont
  {H.}~\bibnamefont {Hafermann}},\ }\href {\doibase 10.1103/PhysRevB.90.235135}
  {\bibfield  {journal} {\bibinfo  {journal} {Phys. Rev. B}\ }\textbf {\bibinfo
  {volume} {90}},\ \bibinfo {pages} {235135} (\bibinfo {year}
  {2014}{\natexlab{a}})}\BibitemShut {NoStop}%
\bibitem [{\citenamefont {Hafermann}\ \emph {et~al.}(2014)\citenamefont
  {Hafermann}, \citenamefont {van Loon}, \citenamefont {Katsnelson},
  \citenamefont {Lichtenstein},\ and\ \citenamefont
  {Parcollet}}]{Hafermann14b}%
  \BibitemOpen
  \bibfield  {author} {\bibinfo {author} {\bibfnamefont {H.}~\bibnamefont
  {Hafermann}}, \bibinfo {author} {\bibfnamefont {E.~G. C.~P.}\ \bibnamefont
  {van Loon}}, \bibinfo {author} {\bibfnamefont {M.~I.}\ \bibnamefont
  {Katsnelson}}, \bibinfo {author} {\bibfnamefont {A.~I.}\ \bibnamefont
  {Lichtenstein}}, \ and\ \bibinfo {author} {\bibfnamefont {O.}~\bibnamefont
  {Parcollet}},\ }\href {\doibase 10.1103/PhysRevB.90.235105} {\bibfield
  {journal} {\bibinfo  {journal} {Phys. Rev. B}\ }\textbf {\bibinfo {volume}
  {90}},\ \bibinfo {pages} {235105} (\bibinfo {year} {2014})}\BibitemShut
  {NoStop}%
\bibitem [{\citenamefont {van Loon}\ \emph
  {et~al.}(2016{\natexlab{a}})\citenamefont {van Loon}, \citenamefont
  {Sch\"uler}, \citenamefont {Katsnelson},\ and\ \citenamefont
  {Wehling}}]{PhysRevB.94.165141}%
  \BibitemOpen
  \bibfield  {author} {\bibinfo {author} {\bibfnamefont {E.~G. C.~P.}\
  \bibnamefont {van Loon}}, \bibinfo {author} {\bibfnamefont {M.}~\bibnamefont
  {Sch\"uler}}, \bibinfo {author} {\bibfnamefont {M.~I.}\ \bibnamefont
  {Katsnelson}}, \ and\ \bibinfo {author} {\bibfnamefont {T.~O.}\ \bibnamefont
  {Wehling}},\ }\href {\doibase 10.1103/PhysRevB.94.165141} {\bibfield
  {journal} {\bibinfo  {journal} {Phys. Rev. B}\ }\textbf {\bibinfo {volume}
  {94}},\ \bibinfo {pages} {165141} (\bibinfo {year}
  {2016}{\natexlab{a}})}\BibitemShut {NoStop}%
\bibitem [{\citenamefont {Terletska}\ \emph {et~al.}(2017)\citenamefont
  {Terletska}, \citenamefont {Chen},\ and\ \citenamefont
  {Gull}}]{PhysRevB.95.115149}%
  \BibitemOpen
  \bibfield  {author} {\bibinfo {author} {\bibfnamefont {H.}~\bibnamefont
  {Terletska}}, \bibinfo {author} {\bibfnamefont {T.}~\bibnamefont {Chen}}, \
  and\ \bibinfo {author} {\bibfnamefont {E.}~\bibnamefont {Gull}},\ }\href
  {\doibase 10.1103/PhysRevB.95.115149} {\bibfield  {journal} {\bibinfo
  {journal} {Phys. Rev. B}\ }\textbf {\bibinfo {volume} {95}},\ \bibinfo
  {pages} {115149} (\bibinfo {year} {2017})}\BibitemShut {NoStop}%
\bibitem [{\citenamefont {Stepanov}\ \emph {et~al.}(2019)\citenamefont
  {Stepanov}, \citenamefont {Huber}, \citenamefont {Lichtenstein},\ and\
  \citenamefont {Katsnelson}}]{PhysRevB.99.115124}%
  \BibitemOpen
  \bibfield  {author} {\bibinfo {author} {\bibfnamefont {E.~A.}\ \bibnamefont
  {Stepanov}}, \bibinfo {author} {\bibfnamefont {A.}~\bibnamefont {Huber}},
  \bibinfo {author} {\bibfnamefont {A.~I.}\ \bibnamefont {Lichtenstein}}, \
  and\ \bibinfo {author} {\bibfnamefont {M.~I.}\ \bibnamefont {Katsnelson}},\
  }\href {\doibase 10.1103/PhysRevB.99.115124} {\bibfield  {journal} {\bibinfo
  {journal} {Phys. Rev. B}\ }\textbf {\bibinfo {volume} {99}},\ \bibinfo
  {pages} {115124} (\bibinfo {year} {2019})}\BibitemShut {NoStop}%
\bibitem [{Note1()}]{Note1}%
  \BibitemOpen
  \bibinfo {note} {The equal time component is obtained by dividing both sides
  by $\beta $ to normalize the frequency sum, and the sum runs over both
  positive and negative Matsubara frequencies.}\BibitemShut {Stop}%
\bibitem [{\citenamefont {van Loon}\ \emph
  {et~al.}(2016{\natexlab{b}})\citenamefont {van Loon}, \citenamefont {Krien},
  \citenamefont {Hafermann}, \citenamefont {Stepanov}, \citenamefont
  {Lichtenstein},\ and\ \citenamefont {Katsnelson}}]{vanLoon16}%
  \BibitemOpen
  \bibfield  {author} {\bibinfo {author} {\bibfnamefont {E.~G. C.~P.}\
  \bibnamefont {van Loon}}, \bibinfo {author} {\bibfnamefont {F.}~\bibnamefont
  {Krien}}, \bibinfo {author} {\bibfnamefont {H.}~\bibnamefont {Hafermann}},
  \bibinfo {author} {\bibfnamefont {E.~A.}\ \bibnamefont {Stepanov}}, \bibinfo
  {author} {\bibfnamefont {A.~I.}\ \bibnamefont {Lichtenstein}}, \ and\
  \bibinfo {author} {\bibfnamefont {M.~I.}\ \bibnamefont {Katsnelson}},\ }\href
  {\doibase 10.1103/PhysRevB.93.155162} {\bibfield  {journal} {\bibinfo
  {journal} {Phys. Rev. B}\ }\textbf {\bibinfo {volume} {93}},\ \bibinfo
  {pages} {155162} (\bibinfo {year} {2016}{\natexlab{b}})}\BibitemShut
  {NoStop}%
\bibitem [{\citenamefont {Baier}\ \emph {et~al.}(2000)\citenamefont {Baier},
  \citenamefont {Bick},\ and\ \citenamefont {Wetterich}}]{Baier00}%
  \BibitemOpen
  \bibfield  {author} {\bibinfo {author} {\bibfnamefont {T.}~\bibnamefont
  {Baier}}, \bibinfo {author} {\bibfnamefont {E.}~\bibnamefont {Bick}}, \ and\
  \bibinfo {author} {\bibfnamefont {C.}~\bibnamefont {Wetterich}},\ }\href
  {\doibase 10.1103/PhysRevB.62.15471} {\bibfield  {journal} {\bibinfo
  {journal} {Phys. Rev. B}\ }\textbf {\bibinfo {volume} {62}},\ \bibinfo
  {pages} {15471} (\bibinfo {year} {2000})}\BibitemShut {NoStop}%
\bibitem [{\citenamefont {Jaeckel}\ and\ \citenamefont
  {Wetterich}(2003)}]{Jaeckel03}%
  \BibitemOpen
  \bibfield  {author} {\bibinfo {author} {\bibfnamefont {J.}~\bibnamefont
  {Jaeckel}}\ and\ \bibinfo {author} {\bibfnamefont {C.}~\bibnamefont
  {Wetterich}},\ }\href {\doibase 10.1103/PhysRevD.68.025020} {\bibfield
  {journal} {\bibinfo  {journal} {Phys. Rev. D}\ }\textbf {\bibinfo {volume}
  {68}},\ \bibinfo {pages} {025020} (\bibinfo {year} {2003})}\BibitemShut
  {NoStop}%
\bibitem [{\citenamefont {Ayral}\ \emph {et~al.}(2017)\citenamefont {Ayral},
  \citenamefont {Vu\ifmmode \check{c}\else \v{c}\fi{}i\ifmmode \check{c}\else
  \v{c}\fi{}evi\ifmmode~\acute{c}\else \'{c}\fi{}},\ and\ \citenamefont
  {Parcollet}}]{Ayral17}%
  \BibitemOpen
  \bibfield  {author} {\bibinfo {author} {\bibfnamefont {T.}~\bibnamefont
  {Ayral}}, \bibinfo {author} {\bibfnamefont {J.}~\bibnamefont {Vu\ifmmode
  \check{c}\else \v{c}\fi{}i\ifmmode \check{c}\else
  \v{c}\fi{}evi\ifmmode~\acute{c}\else \'{c}\fi{}}}, \ and\ \bibinfo {author}
  {\bibfnamefont {O.}~\bibnamefont {Parcollet}},\ }\href {\doibase
  10.1103/PhysRevLett.119.166401} {\bibfield  {journal} {\bibinfo  {journal}
  {Phys. Rev. Lett.}\ }\textbf {\bibinfo {volume} {119}},\ \bibinfo {pages}
  {166401} (\bibinfo {year} {2017})}\BibitemShut {NoStop}%
\bibitem [{\citenamefont {{Del Re}}\ \emph {et~al.}(2018)\citenamefont {{Del
  Re}}, \citenamefont {{Capone}},\ and\ \citenamefont {{Toschi}}}]{delRe18}%
  \BibitemOpen
  \bibfield  {author} {\bibinfo {author} {\bibfnamefont {L.}~\bibnamefont {{Del
  Re}}}, \bibinfo {author} {\bibfnamefont {M.}~\bibnamefont {{Capone}}}, \ and\
  \bibinfo {author} {\bibfnamefont {A.}~\bibnamefont {{Toschi}}},\ }\href@noop
  {} {\bibfield  {journal} {\bibinfo  {journal} {ArXiv e-prints}\ } (\bibinfo
  {year} {2018})},\ \Eprint {http://arxiv.org/abs/1805.05194} {arXiv:1805.05194
  [cond-mat.str-el]} \BibitemShut {NoStop}%
\bibitem [{\citenamefont {Iskakov}\ \emph {et~al.}(2016)\citenamefont
  {Iskakov}, \citenamefont {Antipov},\ and\ \citenamefont {Gull}}]{Iskakov16}%
  \BibitemOpen
  \bibfield  {author} {\bibinfo {author} {\bibfnamefont {S.}~\bibnamefont
  {Iskakov}}, \bibinfo {author} {\bibfnamefont {A.~E.}\ \bibnamefont
  {Antipov}}, \ and\ \bibinfo {author} {\bibfnamefont {E.}~\bibnamefont
  {Gull}},\ }\href {\doibase 10.1103/PhysRevB.94.035102} {\bibfield  {journal}
  {\bibinfo  {journal} {Phys. Rev. B}\ }\textbf {\bibinfo {volume} {94}},\
  \bibinfo {pages} {035102} (\bibinfo {year} {2016})}\BibitemShut {NoStop}%
\bibitem [{\citenamefont {Gukelberger}\ \emph {et~al.}(2017)\citenamefont
  {Gukelberger}, \citenamefont {Kozik},\ and\ \citenamefont
  {Hafermann}}]{Gukelberger2017}%
  \BibitemOpen
  \bibfield  {author} {\bibinfo {author} {\bibfnamefont {J.}~\bibnamefont
  {Gukelberger}}, \bibinfo {author} {\bibfnamefont {E.}~\bibnamefont {Kozik}},
  \ and\ \bibinfo {author} {\bibfnamefont {H.}~\bibnamefont {Hafermann}},\
  }\href {\doibase 10.1103/PhysRevB.96.035152} {\bibfield  {journal} {\bibinfo
  {journal} {Phys. Rev. B}\ }\textbf {\bibinfo {volume} {96}},\ \bibinfo
  {pages} {035152} (\bibinfo {year} {2017})}\BibitemShut {NoStop}%
\bibitem [{\citenamefont {Hafermann}\ \emph {et~al.}(2013)\citenamefont
  {Hafermann}, \citenamefont {Werner},\ and\ \citenamefont
  {Gull}}]{Hafermann13}%
  \BibitemOpen
  \bibfield  {author} {\bibinfo {author} {\bibfnamefont {H.}~\bibnamefont
  {Hafermann}}, \bibinfo {author} {\bibfnamefont {P.}~\bibnamefont {Werner}}, \
  and\ \bibinfo {author} {\bibfnamefont {E.}~\bibnamefont {Gull}},\ }\href
  {\doibase http://dx.doi.org/10.1016/j.cpc.2012.12.013} {\bibfield  {journal}
  {\bibinfo  {journal} {Computer Physics Communications}\ }\textbf {\bibinfo
  {volume} {184}},\ \bibinfo {pages} {1280 } (\bibinfo {year}
  {2013})}\BibitemShut {NoStop}%
\bibitem [{\citenamefont {Hafermann}(2014)}]{Hafermann14}%
  \BibitemOpen
  \bibfield  {author} {\bibinfo {author} {\bibfnamefont {H.}~\bibnamefont
  {Hafermann}},\ }\href {\doibase 10.1103/PhysRevB.89.235128} {\bibfield
  {journal} {\bibinfo  {journal} {Phys. Rev. B}\ }\textbf {\bibinfo {volume}
  {89}},\ \bibinfo {pages} {235128} (\bibinfo {year} {2014})}\BibitemShut
  {NoStop}%
\bibitem [{\citenamefont {Bauer}\ \emph {et~al.}(2011)\citenamefont {Bauer},
  \citenamefont {Carr}, \citenamefont {Evertz}, \citenamefont {Feiguin},
  \citenamefont {Freire}, \citenamefont {Fuchs}, \citenamefont {Gamper},
  \citenamefont {Gukelberger}, \citenamefont {Gull}, \citenamefont {Guertler},
  \citenamefont {Hehn}, \citenamefont {Igarashi}, \citenamefont {Isakov},
  \citenamefont {Koop}, \citenamefont {Ma}, \citenamefont {Mates},
  \citenamefont {Matsuo}, \citenamefont {Parcollet}, \citenamefont
  {Pawłowski}, \citenamefont {Picon}, \citenamefont {Pollet}, \citenamefont
  {Santos}, \citenamefont {Scarola}, \citenamefont {Schollwöck}, \citenamefont
  {Silva}, \citenamefont {Surer}, \citenamefont {Todo}, \citenamefont {Trebst},
  \citenamefont {Troyer}, \citenamefont {Wall}, \citenamefont {Werner},\ and\
  \citenamefont {Wessel}}]{ALPS2}%
  \BibitemOpen
  \bibfield  {author} {\bibinfo {author} {\bibfnamefont {B.}~\bibnamefont
  {Bauer}}, \bibinfo {author} {\bibfnamefont {L.~D.}\ \bibnamefont {Carr}},
  \bibinfo {author} {\bibfnamefont {H.~G.}\ \bibnamefont {Evertz}}, \bibinfo
  {author} {\bibfnamefont {A.}~\bibnamefont {Feiguin}}, \bibinfo {author}
  {\bibfnamefont {J.}~\bibnamefont {Freire}}, \bibinfo {author} {\bibfnamefont
  {S.}~\bibnamefont {Fuchs}}, \bibinfo {author} {\bibfnamefont
  {L.}~\bibnamefont {Gamper}}, \bibinfo {author} {\bibfnamefont
  {J.}~\bibnamefont {Gukelberger}}, \bibinfo {author} {\bibfnamefont
  {E.}~\bibnamefont {Gull}}, \bibinfo {author} {\bibfnamefont {S.}~\bibnamefont
  {Guertler}}, \bibinfo {author} {\bibfnamefont {A.}~\bibnamefont {Hehn}},
  \bibinfo {author} {\bibfnamefont {R.}~\bibnamefont {Igarashi}}, \bibinfo
  {author} {\bibfnamefont {S.~V.}\ \bibnamefont {Isakov}}, \bibinfo {author}
  {\bibfnamefont {D.}~\bibnamefont {Koop}}, \bibinfo {author} {\bibfnamefont
  {P.~N.}\ \bibnamefont {Ma}}, \bibinfo {author} {\bibfnamefont
  {P.}~\bibnamefont {Mates}}, \bibinfo {author} {\bibfnamefont
  {H.}~\bibnamefont {Matsuo}}, \bibinfo {author} {\bibfnamefont
  {O.}~\bibnamefont {Parcollet}}, \bibinfo {author} {\bibfnamefont
  {G.}~\bibnamefont {Pawłowski}}, \bibinfo {author} {\bibfnamefont {J.~D.}\
  \bibnamefont {Picon}}, \bibinfo {author} {\bibfnamefont {L.}~\bibnamefont
  {Pollet}}, \bibinfo {author} {\bibfnamefont {E.}~\bibnamefont {Santos}},
  \bibinfo {author} {\bibfnamefont {V.~W.}\ \bibnamefont {Scarola}}, \bibinfo
  {author} {\bibfnamefont {U.}~\bibnamefont {Schollwöck}}, \bibinfo {author}
  {\bibfnamefont {C.}~\bibnamefont {Silva}}, \bibinfo {author} {\bibfnamefont
  {B.}~\bibnamefont {Surer}}, \bibinfo {author} {\bibfnamefont
  {S.}~\bibnamefont {Todo}}, \bibinfo {author} {\bibfnamefont {S.}~\bibnamefont
  {Trebst}}, \bibinfo {author} {\bibfnamefont {M.}~\bibnamefont {Troyer}},
  \bibinfo {author} {\bibfnamefont {M.~L.}\ \bibnamefont {Wall}}, \bibinfo
  {author} {\bibfnamefont {P.}~\bibnamefont {Werner}}, \ and\ \bibinfo {author}
  {\bibfnamefont {S.}~\bibnamefont {Wessel}},\ }\href {\doibase
  10.1088/1742-5468/2011/05/P05001} {\bibfield  {journal} {\bibinfo  {journal}
  {Journal of Statistical Mechanics: Theory and Experiment}\ }\textbf {\bibinfo
  {volume} {2011}},\ \bibinfo {pages} {P05001} (\bibinfo {year}
  {2011})}\BibitemShut {NoStop}%
\bibitem [{Note2()}]{Note2}%
  \BibitemOpen
  \bibinfo {note} {The sign of $X$ is fixed, so there can be no cancellation of
  divergences on the right hand side.}\BibitemShut {Stop}%
\bibitem [{\citenamefont {Maier}\ \emph {et~al.}(2005)\citenamefont {Maier},
  \citenamefont {Jarrell}, \citenamefont {Pruschke},\ and\ \citenamefont
  {Hettler}}]{Maier05}%
  \BibitemOpen
  \bibfield  {author} {\bibinfo {author} {\bibfnamefont {T.}~\bibnamefont
  {Maier}}, \bibinfo {author} {\bibfnamefont {M.}~\bibnamefont {Jarrell}},
  \bibinfo {author} {\bibfnamefont {T.}~\bibnamefont {Pruschke}}, \ and\
  \bibinfo {author} {\bibfnamefont {M.~H.}\ \bibnamefont {Hettler}},\ }\href
  {\doibase 10.1103/RevModPhys.77.1027} {\bibfield  {journal} {\bibinfo
  {journal} {Rev. Mod. Phys.}\ }\textbf {\bibinfo {volume} {77}},\ \bibinfo
  {pages} {1027} (\bibinfo {year} {2005})}\BibitemShut {NoStop}%
\bibitem [{\citenamefont {Mermin}\ and\ \citenamefont
  {Wagner}(1966)}]{Mermin66}%
  \BibitemOpen
  \bibfield  {author} {\bibinfo {author} {\bibfnamefont {N.~D.}\ \bibnamefont
  {Mermin}}\ and\ \bibinfo {author} {\bibfnamefont {H.}~\bibnamefont
  {Wagner}},\ }\href {\doibase 10.1103/PhysRevLett.17.1133} {\bibfield
  {journal} {\bibinfo  {journal} {Phys. Rev. Lett.}\ }\textbf {\bibinfo
  {volume} {17}},\ \bibinfo {pages} {1133} (\bibinfo {year}
  {1966})}\BibitemShut {NoStop}%
\bibitem [{\citenamefont {Irkhin}\ \emph {et~al.}(1999)\citenamefont {Irkhin},
  \citenamefont {Katanin},\ and\ \citenamefont {Katsnelson}}]{IKK}%
  \BibitemOpen
  \bibfield  {author} {\bibinfo {author} {\bibfnamefont {V.~Y.}\ \bibnamefont
  {Irkhin}}, \bibinfo {author} {\bibfnamefont {A.~A.}\ \bibnamefont {Katanin}},
  \ and\ \bibinfo {author} {\bibfnamefont {M.~I.}\ \bibnamefont {Katsnelson}},\
  }\href@noop {} {\bibfield  {journal} {\bibinfo  {journal} {Phys. Rev. B}\
  }\textbf {\bibinfo {volume} {60}},\ \bibinfo {pages} {1082} (\bibinfo {year}
  {1999})}\BibitemShut {NoStop}%
\bibitem [{\citenamefont {Chakravarty}\ \emph {et~al.}(1989)\citenamefont
  {Chakravarty}, \citenamefont {Halperin},\ and\ \citenamefont
  {Nelson}}]{PhysRevB.39.2344}%
  \BibitemOpen
  \bibfield  {author} {\bibinfo {author} {\bibfnamefont {S.}~\bibnamefont
  {Chakravarty}}, \bibinfo {author} {\bibfnamefont {B.~I.}\ \bibnamefont
  {Halperin}}, \ and\ \bibinfo {author} {\bibfnamefont {D.~R.}\ \bibnamefont
  {Nelson}},\ }\href {\doibase 10.1103/PhysRevB.39.2344} {\bibfield  {journal}
  {\bibinfo  {journal} {Phys. Rev. B}\ }\textbf {\bibinfo {volume} {39}},\
  \bibinfo {pages} {2344} (\bibinfo {year} {1989})}\BibitemShut {NoStop}%
\bibitem [{Note3()}]{Note3}%
  \BibitemOpen
  \bibinfo {note} {It is worth observing that these relations occur on very
  different length and time scales: The Mermin-Wagner theorem and Goldstone
  modes are long wavelength, low frequency phenomena whereas the high-frequency
  asymptote of the self-energy is a local, high frequency
  phenomenon.}\BibitemShut {Stop}%
\bibitem [{\citenamefont {Baym}\ and\ \citenamefont {Kadanoff}(1961)}]{Baym61}%
  \BibitemOpen
  \bibfield  {author} {\bibinfo {author} {\bibfnamefont {G.}~\bibnamefont
  {Baym}}\ and\ \bibinfo {author} {\bibfnamefont {L.~P.}\ \bibnamefont
  {Kadanoff}},\ }\href {\doibase 10.1103/PhysRev.124.287} {\bibfield  {journal}
  {\bibinfo  {journal} {Phys. Rev.}\ }\textbf {\bibinfo {volume} {124}},\
  \bibinfo {pages} {287} (\bibinfo {year} {1961})}\BibitemShut {NoStop}%
\bibitem [{\citenamefont {Baym}(1962)}]{Baym62}%
  \BibitemOpen
  \bibfield  {author} {\bibinfo {author} {\bibfnamefont {G.}~\bibnamefont
  {Baym}},\ }\href {\doibase 10.1103/PhysRev.127.1391} {\bibfield  {journal}
  {\bibinfo  {journal} {Phys. Rev.}\ }\textbf {\bibinfo {volume} {127}},\
  \bibinfo {pages} {1391} (\bibinfo {year} {1962})}\BibitemShut {NoStop}%
\bibitem [{\citenamefont {Geldart}\ and\ \citenamefont
  {Taylor}(1970)}]{Geldart70}%
  \BibitemOpen
  \bibfield  {author} {\bibinfo {author} {\bibfnamefont {D.~J.~W.}\
  \bibnamefont {Geldart}}\ and\ \bibinfo {author} {\bibfnamefont
  {R.}~\bibnamefont {Taylor}},\ }\href {\doibase 10.1139/p70-022} {\bibfield
  {journal} {\bibinfo  {journal} {Canadian Journal of Physics}\ }\textbf
  {\bibinfo {volume} {48}},\ \bibinfo {pages} {155} (\bibinfo {year} {1970})},\
  \Eprint {http://arxiv.org/abs/http://dx.doi.org/10.1139/p70-022}
  {http://dx.doi.org/10.1139/p70-022} \BibitemShut {NoStop}%
\bibitem [{\citenamefont {Moriya}\ and\ \citenamefont
  {Kawabata}(1973)}]{Moriya1973}%
  \BibitemOpen
  \bibfield  {author} {\bibinfo {author} {\bibfnamefont {T.}~\bibnamefont
  {Moriya}}\ and\ \bibinfo {author} {\bibfnamefont {A.}~\bibnamefont
  {Kawabata}},\ }\href {\doibase 10.1143/JPSJ.34.639} {\bibfield  {journal}
  {\bibinfo  {journal} {Journal of the Physical Society of Japan}\ }\textbf
  {\bibinfo {volume} {34}},\ \bibinfo {pages} {639} (\bibinfo {year} {1973})},\
  \Eprint {http://arxiv.org/abs/http://dx.doi.org/10.1143/JPSJ.34.639}
  {http://dx.doi.org/10.1143/JPSJ.34.639} \BibitemShut {NoStop}%
\bibitem [{\citenamefont {Dzyaloshinskii}\ and\ \citenamefont
  {Kondratenko}(1976)}]{Dzyaloshinskii1976}%
  \BibitemOpen
  \bibfield  {author} {\bibinfo {author} {\bibfnamefont {I.}~\bibnamefont
  {Dzyaloshinskii}}\ and\ \bibinfo {author} {\bibfnamefont {P.}~\bibnamefont
  {Kondratenko}},\ }\href@noop {} {\bibfield  {journal} {\bibinfo  {journal}
  {Soviet Journal of Experimental and Theoretical Physics}\ }\textbf {\bibinfo
  {volume} {43}},\ \bibinfo {pages} {1036} (\bibinfo {year}
  {1976})}\BibitemShut {NoStop}%
\bibitem [{\citenamefont {Moriya}(1985)}]{Moriya1985}%
  \BibitemOpen
  \bibfield  {author} {\bibinfo {author} {\bibfnamefont {T.}~\bibnamefont
  {Moriya}},\ }\href@noop {} {\emph {\bibinfo {title} {Spin fluctuations in
  itinerant electron magnetism}}},\ Vol.~\bibinfo {volume} {56}\ (\bibinfo
  {publisher} {Springer-Verlag Berlin},\ \bibinfo {year} {1985})\BibitemShut
  {NoStop}%
\bibitem [{\citenamefont {Aichhorn}\ \emph {et~al.}(2006)\citenamefont
  {Aichhorn}, \citenamefont {Arrigoni}, \citenamefont {Potthoff},\ and\
  \citenamefont {Hanke}}]{Aichhorn06}%
  \BibitemOpen
  \bibfield  {author} {\bibinfo {author} {\bibfnamefont {M.}~\bibnamefont
  {Aichhorn}}, \bibinfo {author} {\bibfnamefont {E.}~\bibnamefont {Arrigoni}},
  \bibinfo {author} {\bibfnamefont {M.}~\bibnamefont {Potthoff}}, \ and\
  \bibinfo {author} {\bibfnamefont {W.}~\bibnamefont {Hanke}},\ }\href
  {\doibase 10.1103/PhysRevB.74.024508} {\bibfield  {journal} {\bibinfo
  {journal} {Phys. Rev. B}\ }\textbf {\bibinfo {volume} {74}},\ \bibinfo
  {pages} {024508} (\bibinfo {year} {2006})}\BibitemShut {NoStop}%
\bibitem [{\citenamefont {Potthoff}(2006)}]{Potthoff06}%
  \BibitemOpen
  \bibfield  {author} {\bibinfo {author} {\bibfnamefont {M.}~\bibnamefont
  {Potthoff}},\ }in\ \href {\doibase 10.1007/11423256_11} {{\selectlanguage
  {English}\emph {\bibinfo {booktitle} {Advances in Solid State Physics}}}},\
  \bibinfo {series} {Advances in Solid State Physics}, Vol.~\bibinfo {volume}
  {45},\ \bibinfo {editor} {edited by\ \bibinfo {editor} {\bibfnamefont
  {B.}~\bibnamefont {Kramer}}}\ (\bibinfo  {publisher} {Springer Berlin
  Heidelberg},\ \bibinfo {year} {2006})\ pp.\ \bibinfo {pages}
  {135--147}\BibitemShut {NoStop}%
\bibitem [{\citenamefont {Fishman}\ \emph {et~al.}(2005)\citenamefont
  {Fishman}, \citenamefont {Moreno}, \citenamefont {Maier},\ and\ \citenamefont
  {Jarrell}}]{Fishman05}%
  \BibitemOpen
  \bibfield  {author} {\bibinfo {author} {\bibfnamefont {R.~S.}\ \bibnamefont
  {Fishman}}, \bibinfo {author} {\bibfnamefont {J.}~\bibnamefont {Moreno}},
  \bibinfo {author} {\bibfnamefont {T.}~\bibnamefont {Maier}}, \ and\ \bibinfo
  {author} {\bibfnamefont {M.}~\bibnamefont {Jarrell}},\ }\href {\doibase
  10.1103/PhysRevB.71.180405} {\bibfield  {journal} {\bibinfo  {journal} {Phys.
  Rev. B}\ }\textbf {\bibinfo {volume} {71}},\ \bibinfo {pages} {180405(R)}
  (\bibinfo {year} {2005})}\BibitemShut {NoStop}%
\bibitem [{\citenamefont {Otsuki}\ and\ \citenamefont
  {Kuramoto}(2013)}]{Otsuki13b}%
  \BibitemOpen
  \bibfield  {author} {\bibinfo {author} {\bibfnamefont {J.}~\bibnamefont
  {Otsuki}}\ and\ \bibinfo {author} {\bibfnamefont {Y.}~\bibnamefont
  {Kuramoto}},\ }\href {\doibase 10.1103/PhysRevB.88.024427} {\bibfield
  {journal} {\bibinfo  {journal} {Phys. Rev. B}\ }\textbf {\bibinfo {volume}
  {88}},\ \bibinfo {pages} {024427} (\bibinfo {year} {2013})}\BibitemShut
  {NoStop}%
\bibitem [{\citenamefont {van Loon}\ \emph {et~al.}(2015)\citenamefont {van
  Loon}, \citenamefont {Hafermann}, \citenamefont {Lichtenstein},\ and\
  \citenamefont {Katsnelson}}]{vanLoon15}%
  \BibitemOpen
  \bibfield  {author} {\bibinfo {author} {\bibfnamefont {E.~G. C.~P.}\
  \bibnamefont {van Loon}}, \bibinfo {author} {\bibfnamefont {H.}~\bibnamefont
  {Hafermann}}, \bibinfo {author} {\bibfnamefont {A.~I.}\ \bibnamefont
  {Lichtenstein}}, \ and\ \bibinfo {author} {\bibfnamefont {M.~I.}\
  \bibnamefont {Katsnelson}},\ }\href {\doibase 10.1103/PhysRevB.92.085106}
  {\bibfield  {journal} {\bibinfo  {journal} {Phys. Rev. B}\ }\textbf {\bibinfo
  {volume} {92}},\ \bibinfo {pages} {085106} (\bibinfo {year}
  {2015})}\BibitemShut {NoStop}%
\bibitem [{\citenamefont {van Loon}\ \emph
  {et~al.}(2014{\natexlab{b}})\citenamefont {van Loon}, \citenamefont
  {Hafermann}, \citenamefont {Lichtenstein}, \citenamefont {Rubtsov},\ and\
  \citenamefont {Katsnelson}}]{vanLoon14b}%
  \BibitemOpen
  \bibfield  {author} {\bibinfo {author} {\bibfnamefont {E.~G. C.~P.}\
  \bibnamefont {van Loon}}, \bibinfo {author} {\bibfnamefont {H.}~\bibnamefont
  {Hafermann}}, \bibinfo {author} {\bibfnamefont {A.~I.}\ \bibnamefont
  {Lichtenstein}}, \bibinfo {author} {\bibfnamefont {A.~N.}\ \bibnamefont
  {Rubtsov}}, \ and\ \bibinfo {author} {\bibfnamefont {M.~I.}\ \bibnamefont
  {Katsnelson}},\ }\href {\doibase 10.1103/PhysRevLett.113.246407} {\bibfield
  {journal} {\bibinfo  {journal} {Phys. Rev. Lett.}\ }\textbf {\bibinfo
  {volume} {113}},\ \bibinfo {pages} {246407} (\bibinfo {year}
  {2014}{\natexlab{b}})}\BibitemShut {NoStop}%
\bibitem [{\citenamefont {{Geffroy}}\ \emph {et~al.}(2018)\citenamefont
  {{Geffroy}}, \citenamefont {{Kaufmann}}, \citenamefont {{Hariki}},
  \citenamefont {{Gunacker}}, \citenamefont {{Hausoel}},\ and\ \citenamefont
  {{Kunes}}}]{Geffroy18}%
  \BibitemOpen
  \bibfield  {author} {\bibinfo {author} {\bibfnamefont {D.}~\bibnamefont
  {{Geffroy}}}, \bibinfo {author} {\bibfnamefont {J.}~\bibnamefont
  {{Kaufmann}}}, \bibinfo {author} {\bibfnamefont {A.}~\bibnamefont
  {{Hariki}}}, \bibinfo {author} {\bibfnamefont {P.}~\bibnamefont
  {{Gunacker}}}, \bibinfo {author} {\bibfnamefont {A.}~\bibnamefont
  {{Hausoel}}}, \ and\ \bibinfo {author} {\bibfnamefont {J.}~\bibnamefont
  {{Kunes}}},\ }\href@noop {} {\bibfield  {journal} {\bibinfo  {journal} {ArXiv
  e-prints}\ } (\bibinfo {year} {2018})},\ \Eprint
  {http://arxiv.org/abs/1808.08046} {arXiv:1808.08046 [cond-mat.str-el]}
  \BibitemShut {NoStop}%
\bibitem [{Note4()}]{Note4}%
  \BibitemOpen
  \bibinfo {note} {See, e.g., the review~\protect \rev@citealpnum
  {Rohringer18}.}\BibitemShut {Stop}%
\bibitem [{\citenamefont {Rohringer}\ and\ \citenamefont
  {Toschi}(2016)}]{Rohringer16}%
  \BibitemOpen
  \bibfield  {author} {\bibinfo {author} {\bibfnamefont {G.}~\bibnamefont
  {Rohringer}}\ and\ \bibinfo {author} {\bibfnamefont {A.}~\bibnamefont
  {Toschi}},\ }\href {\doibase 10.1103/PhysRevB.94.125144} {\bibfield
  {journal} {\bibinfo  {journal} {Phys. Rev. B}\ }\textbf {\bibinfo {volume}
  {94}},\ \bibinfo {pages} {125144} (\bibinfo {year} {2016})}\BibitemShut
  {NoStop}%
\bibitem [{\citenamefont {{Esterling}}(2018{\natexlab{a}})}]{Esterling18}%
  \BibitemOpen
  \bibfield  {author} {\bibinfo {author} {\bibfnamefont {D.~M.}\ \bibnamefont
  {{Esterling}}},\ }\href@noop {} {\bibfield  {journal} {\bibinfo  {journal}
  {ArXiv e-prints}\ } (\bibinfo {year} {2018}{\natexlab{a}})},\ \Eprint
  {http://arxiv.org/abs/1807.09219} {arXiv:1807.09219 [cond-mat.str-el]}
  \BibitemShut {NoStop}%
\bibitem [{\citenamefont {{Esterling}}(2018{\natexlab{b}})}]{Esterling18b}%
  \BibitemOpen
  \bibfield  {author} {\bibinfo {author} {\bibfnamefont {D.~M.}\ \bibnamefont
  {{Esterling}}},\ }\href@noop {} {\bibfield  {journal} {\bibinfo  {journal}
  {ArXiv e-prints}\ } (\bibinfo {year} {2018}{\natexlab{b}})},\ \Eprint
  {http://arxiv.org/abs/1807.09223} {arXiv:1807.09223 [cond-mat.str-el]}
  \BibitemShut {NoStop}%
\bibitem [{\citenamefont {Katanin}\ \emph {et~al.}(2009)\citenamefont
  {Katanin}, \citenamefont {Toschi},\ and\ \citenamefont {Held}}]{Katanin2009}%
  \BibitemOpen
  \bibfield  {author} {\bibinfo {author} {\bibfnamefont {A.~A.}\ \bibnamefont
  {Katanin}}, \bibinfo {author} {\bibfnamefont {A.}~\bibnamefont {Toschi}}, \
  and\ \bibinfo {author} {\bibfnamefont {K.}~\bibnamefont {Held}},\ }\href
  {\doibase 10.1103/PhysRevB.80.075104} {\bibfield  {journal} {\bibinfo
  {journal} {Phys. Rev. B}\ }\textbf {\bibinfo {volume} {80}},\ \bibinfo
  {pages} {075104} (\bibinfo {year} {2009})}\BibitemShut {NoStop}%
\bibitem [{\citenamefont {Parcollet}\ \emph {et~al.}(2015)\citenamefont
  {Parcollet}, \citenamefont {Ferrero}, \citenamefont {Ayral}, \citenamefont
  {Hafermann}, \citenamefont {Krivenko}, \citenamefont {Messio},\ and\
  \citenamefont {Seth}}]{triqs}%
  \BibitemOpen
  \bibfield  {author} {\bibinfo {author} {\bibfnamefont {O.}~\bibnamefont
  {Parcollet}}, \bibinfo {author} {\bibfnamefont {M.}~\bibnamefont {Ferrero}},
  \bibinfo {author} {\bibfnamefont {T.}~\bibnamefont {Ayral}}, \bibinfo
  {author} {\bibfnamefont {H.}~\bibnamefont {Hafermann}}, \bibinfo {author}
  {\bibfnamefont {I.}~\bibnamefont {Krivenko}}, \bibinfo {author}
  {\bibfnamefont {L.}~\bibnamefont {Messio}}, \ and\ \bibinfo {author}
  {\bibfnamefont {P.}~\bibnamefont {Seth}},\ }\href {\doibase
  http://dx.doi.org/10.1016/j.cpc.2015.04.023} {\bibfield  {journal} {\bibinfo
  {journal} {Computer Physics Communications}\ }\textbf {\bibinfo {volume}
  {196}},\ \bibinfo {pages} {398 } (\bibinfo {year} {2015})}\BibitemShut
  {NoStop}%
\bibitem [{Note5()}]{Note5}%
  \BibitemOpen
  \bibinfo {note} {Note that only positive frequencies are shown, so that all
  finite frequencies have a negative frequency counterpart that is equal by
  symmetry.}\BibitemShut {Stop}%
\bibitem [{\citenamefont {Haule}(2007)}]{Haule07}%
  \BibitemOpen
  \bibfield  {author} {\bibinfo {author} {\bibfnamefont {K.}~\bibnamefont
  {Haule}},\ }\href {\doibase 10.1103/PhysRevB.75.155113} {\bibfield  {journal}
  {\bibinfo  {journal} {Phys. Rev. B}\ }\textbf {\bibinfo {volume} {75}},\
  \bibinfo {pages} {155113} (\bibinfo {year} {2007})}\BibitemShut {NoStop}%
\bibitem [{Note6()}]{Note6}%
  \BibitemOpen
  \bibinfo {note} {Note that we find a slope of roughly $X^{-1} \propto -7V$.
  Without any correlation effects, a slope of $-6V=-V_{q=(\pi ,\pi ,\pi )}$
  would be expected from Eq.~\protect \textup {\hbox {\mathsurround \z@
  \protect \normalfont (\ignorespaces \ref {eq:X}\unskip \@@italiccorr
  )}}.}\BibitemShut {Stop}%
\bibitem [{\citenamefont {Vonsovsky}\ and\ \citenamefont
  {Katsnelson}(1979)}]{Vonsovsky_1979}%
  \BibitemOpen
  \bibfield  {author} {\bibinfo {author} {\bibfnamefont {S.~V.}\ \bibnamefont
  {Vonsovsky}}\ and\ \bibinfo {author} {\bibfnamefont {M.~I.}\ \bibnamefont
  {Katsnelson}},\ }\href {\doibase 10.1088/0022-3719/12/11/015} {\bibfield
  {journal} {\bibinfo  {journal} {Journal of Physics C: Solid State Physics}\
  }\textbf {\bibinfo {volume} {12}},\ \bibinfo {pages} {2043} (\bibinfo {year}
  {1979})}\BibitemShut {NoStop}%
\bibitem [{Note7()}]{Note7}%
  \BibitemOpen
  \bibinfo {note} {For the Hubbard model, $\protect \mathaccentV
  {tilde}07E{X}^{(0)}$ is always local. The self-consistency condition $\DOTSB
  \sum@ \slimits@ _\protect \mathbf {q}\protect \mathaccentV
  {tilde}07E{X}^{(0)}=0$ then automatically ensures that $\protect \mathaccentV
  {tilde}07E{X}=0$ everywhere and that Dual Boson reduces to Dual Fermion. The
  self-consistency condition used in this work does not require $\DOTSB \sum@
  \slimits@ _\protect \mathbf {q}\protect \mathaccentV {tilde}07E{X}^{(0)}=0$
  and $\protect \mathaccentV {tilde}07E{X}$ is finite in the Hubbard
  model.}\BibitemShut {Stop}%
\bibitem [{\citenamefont {van Loon}\ \emph
  {et~al.}(2018{\natexlab{a}})\citenamefont {van Loon}, \citenamefont {Krien},
  \citenamefont {Hafermann}, \citenamefont {Lichtenstein},\ and\ \citenamefont
  {Katsnelson}}]{vanLoon18}%
  \BibitemOpen
  \bibfield  {author} {\bibinfo {author} {\bibfnamefont {E.~G. C.~P.}\
  \bibnamefont {van Loon}}, \bibinfo {author} {\bibfnamefont {F.}~\bibnamefont
  {Krien}}, \bibinfo {author} {\bibfnamefont {H.}~\bibnamefont {Hafermann}},
  \bibinfo {author} {\bibfnamefont {A.~I.}\ \bibnamefont {Lichtenstein}}, \
  and\ \bibinfo {author} {\bibfnamefont {M.~I.}\ \bibnamefont {Katsnelson}},\
  }\href {\doibase 10.1103/PhysRevB.98.205148} {\bibfield  {journal} {\bibinfo
  {journal} {Phys. Rev. B}\ }\textbf {\bibinfo {volume} {98}},\ \bibinfo
  {pages} {205148} (\bibinfo {year} {2018}{\natexlab{a}})}\BibitemShut
  {NoStop}%
\bibitem [{\citenamefont {van Loon}\ \emph
  {et~al.}(2018{\natexlab{b}})\citenamefont {van Loon}, \citenamefont
  {Katsnelson},\ and\ \citenamefont {Hafermann}}]{vanLoon18b}%
  \BibitemOpen
  \bibfield  {author} {\bibinfo {author} {\bibfnamefont {E.~G. C.~P.}\
  \bibnamefont {van Loon}}, \bibinfo {author} {\bibfnamefont {M.~I.}\
  \bibnamefont {Katsnelson}}, \ and\ \bibinfo {author} {\bibfnamefont
  {H.}~\bibnamefont {Hafermann}},\ }\href {\doibase 10.1103/PhysRevB.98.155117}
  {\bibfield  {journal} {\bibinfo  {journal} {Phys. Rev. B}\ }\textbf {\bibinfo
  {volume} {98}},\ \bibinfo {pages} {155117} (\bibinfo {year}
  {2018}{\natexlab{b}})}\BibitemShut {NoStop}%
\end{thebibliography}%

\end{document}